\documentclass[12pt]{iopart}
\usepackage{graphicx}
\usepackage{subcaption}
\usepackage{iopams}
\expandafter\let\csname equation*\endcsname\relax
\expandafter\let\csname endequation*\endcsname\relax
\usepackage{amsmath}
\usepackage{epstopdf}
\usepackage{breakcites}

\begin{document}
\title{Charge Exchange Losses and Stochastic Acceleration in the Solar Wind}
\author{C. Kenny, P. Duffy}
\address{School of Physics, University College Dublin, Belfield, Dublin 4, Ireland}
\ead{ciaranmkenny@gmail.com}
\begin{abstract}
Stochastic acceleration of particles under a pressure balance condition can accommodate the universal $p^{-5}$ spectra observed under many different conditions in the inner heliosphere. In this model, in order to avoid an infinite build up of particle pressure, a relationship between the momentum diffusion of particles and the adiabatic deceleration in the solar wind must exist. This constrains both the spatial and momentum diffusion coefficients and results in the $p^{-5}$ spectrum in the presence of adiabatic losses in the solar wind. However, this theory cannot explain the presence of such spectra beyond the termination shock, where adiabatic deceleration is negligible. To explain this apparent discrepancy, we include the effect of charge exchange losses, resulting in new forms of both the spatial and momentum diffusion coefficients that have not previously been considered. Assuming that the turbulence is of a large-scale compressible nature, we find that a balance between momentum diffusion and losses can still readily lead to the creation of $p^{-5}$ suprathermal tails, including those found in the outer heliosphere. 
\end{abstract}

\maketitle

\section{Introduction} \label{sec:intro}
Within the heliosphere and beyond, particles with energies above their expected thermal energies, so-called suprathermal particles, are ubiquitous. Data from \textit{ACE} \cite{fisk12} and \textit{Wind} \cite{day09} among others demonstrate that their spectra commonly take a form close to $ f \propto p^{-5}$, where $f$ is the isotropic phase space distribution function and $p$ is the particle momentum. This spectrum is found both in quiet time and disturbed conditions, near and far from shocks, and in the inner and outer heliosphere. This  implies  that  such a spectrum is independent of local plasma conditions, and that a theory which is not sensitive to the local environment is necessary.

As these tail particles are observed both in quiet times and in more extreme conditions \cite{fisk12}, their acceleration is typically attributed to a stochastic process. Various stochastic theories have been considered in the literature as possible explanations for the origin of these tail particles. One of the primary difficulties in any application of a stochastic theory is the treatment of spatial diffusion. 
 In some instances, spatial diffusion is neglected or considered unimportant compared to other transport processes \cite{zhang12}. In other models, spatial diffusion is treated in an atypical manner. For example, in a series of papers by Fisk and Gloeckler \cite{fisk06, fisk07, fisk08, fisk09, fisk10, fisk12, fisk13, fisk14}, a pump mechanism is developed, where tail particles gain their energy from a continuous ``pumping'' of energy from core particles. This approach naturally leads to the creation of $p^{-5}$ spectra; however, it requires approximating spatial diffusion by a loss term of the form $-f/\tau_E$, where $\tau_E$ is the escape time from a compression region. The validity of this approximation has been discussed in the literature (e.g. \cite{jok10}).

Recently, a new approach has been adopted by several authors, a so-called ``pressure balance'' condition \cite{zhang10,zhang12} \cite{ant13}. As particles stochastically accelerate in the presence of turbulence, their bulk pressure increases. As this turbulence is a finite source of energy and particle pressure, the process cannot continue indefinitely. However, if the increase in particle pressure is ``balanced'' by a source of pressure reduction, such as adiabatic deceleration, then momentum diffusion can be sustained. If we assume that underlying processes for the excitation and dissipation of plasma turbulence constrain the relationship between spatial and momentum diffusion in the presence of adiabatic losses, this condition allows us to determine the particle spectrum.

As an example, consider one of the first applications of this pressure balance condition \cite{zhang12}. Here, the authors considered the stochastic acceleration of particles in a bi-modal plasma, consisting of regions of compressible turbulence and particle acceleration, and regions of no turbulence and no acceleration. Neglecting spatial diffusion, but including charge exchange losses, and assuming a momentum diffusion coefficient of the form $D(p)=D_0 p^2$ , the pressure balance condition allows for an estimation of $D_0$. Under many different circumstances, this leads to the creation of momentum power law spectra in both the turbulent and non-turbulent regions with power law indices of $-5$.
 
This pressure balance condition between momentum diffusion and adiabatic cooling was also applied in \cite{ant13}, herein referred to as ASZ2013. For the first time, pressure balance was applied in the presence of spatial diffusion, albeit in the absence of losses. Once again, this resulted in the creation of power law spectra with spectral indices of $-5$ at large momenta. However, as was discussed in ASZ2013, this pressure balance cannot be sustained in the outer heliosphere, where adiabatic cooling is considered negligible. 

While charge exchange losses have been included in \cite{zhang12} and spatial diffusion has been included in \cite{ant13}, both effects under pressure balance have not been considered together in the literature. It is the purpose of this paper to examine the role of both processes on the resulting spectrum and, in particular, the presence of suprathermal tails past the termination shock.

\section{Pressure Balance} \label{sum_ant}
Assuming a constant solar wind speed $V_0$, a suitable spherically symmetric transport equation for particle acceleration in the presence of turbulence is given by
\begin{equation} \label{or_tran}
\frac{\partial f}{\partial t} + V_0 \frac{\partial f}{\partial r} = \frac{2 V_0}{3 r} p \frac{\partial f}{\partial p} + \frac{1}{r^2}\frac{\partial}{\partial r} \left ( r^2 \kappa \frac{\partial f}{\partial r} \right )   + \frac{1}{p^2} \frac{\partial}{\partial p} \left ( p^2 D \frac{\partial f}{\partial p} \right ) +Q - \frac{f}{\tau_L}
\end{equation} 
where $\kappa(p,r)$ and $D(p,r)$ are the spatial and momentum diffusion coefficients respectively, $Q(r,p)$ is a source term and $\tau_L(r,p)$ is the timescale for losses, which we assume to be caused only by charge exchange. If we assume that the turbulence is composed of magnetosonic waves then, in the case of an infinite correlation time, momentum diffusion is maximised when both diffusion coefficients are related by \cite{zhang13}
\begin{equation} \label{eqn:bal_D}
D(p,r)= \frac{V_c^2 p^2}{15\kappa(p,r)}
\end{equation}
where $V_c$ is the compressional wave speed. This form of momentum diffusion coefficient is reasonable as it follows the same $D \propto p^2/\kappa$ dependence as adopted by other authors (e.g. \cite{jok10}, equation 20 therein). 

In a co-moving frame away from possible particle sources and in an environment where both spatial diffusion and losses are considered unimportant, steady state solutions are obtained by solving
\begin{equation} \label{eqn:comove}
 \frac{2 V_0}{3 r} p \frac{\partial f}{\partial p}  + \frac{1}{p^2} \frac{\partial}{\partial p} \left ( p^2 D \frac{\partial f}{\partial p} \right ) = 0 
\end{equation} 
If $D(p,r) = D_0 p^2/r$, i.e. $\kappa(p,r) \propto r$ by equation \eqref{eqn:bal_D}, then this equation has power law solutions of the form $f \propto p^{a}$, where $a = -(3 + 2V_0/3D_0)$. A particular value for $a$ therefore requires motivating the scaling of $\kappa\propto r$ and an implicit relationship between $D_0$ and $V_0$. These can be motivated by considering how the particle pressure evolves with time.

\subsection{Particle Pressure}
The particle pressure $P(r,t)$  is related to the particle distribution function $f(p,r,t)$ by
\begin{equation}
P(r,t) = \frac{4 \pi}{3m} \int p^4 f(p,r,t) dp
\end{equation}
Multiplying equation \eqref{or_tran} by $4 \pi p^4/3m$ and integrating over momentum, we obtain the following pressure equation
\begin{equation}
\frac{\partial P}{\partial t} + V_0 \frac{\partial P}{\partial r} = -\frac{10V_0}{3r}P + \frac{1}{r^2}\frac{\partial}{\partial r} \left ( r^2 \kappa (r) \frac{\partial P}{\partial r} \right )   + \frac{2V_c^2}{3\kappa(r)} P  +\dot{P_0} - \frac{P}{\tau_L(r)}
\end{equation}
where $\dot{P_0}$ is the contribution to the pressure from the injected particles and we have assumed that both the spatial diffusion coefficient and loss timescale are momentum independent. Let us now consider the growth rate of the pressure by inserting a solution of the form $P \propto e^{\gamma t}$. This results in the following equation for $\gamma$
\begin{equation} \label{gam_bal}
\gamma =  -\frac{10V_0}{3r} + \frac{2V_c^2}{3\kappa(r)} - \frac{1}{\tau_L(r)}
\end{equation}
With underlying processes of turbulence excitation and damping by the energetic particles and the background plasma we look for solutions where the pressure does not grow arbitrarily large and where $\gamma=0$ - the ``pressure balance'' condition. Equation 
\label{eqn:gam_bal} then relates the diffusion coefficients to the loss timescale
\begin{align} \label{eqn:bal_kappa}
\kappa(r) = \frac{V_c^2 r}{5V_0 + 3r/2\tau_L} && D(p,r) = \frac{p^2 V_0}{3r} \left(1 + \frac{3r}{10V_0 \tau_L}\right)
\end{align}
where we have used equation \eqref{eqn:bal_D} to obtain the momentum diffusion coefficient. For the case considered in equation \eqref{eqn:comove}, the momentum diffusion coefficient is reduced to the form $D(p,r) = D_0 p^2/r$ where $D_0 = V_0/3$. This coefficient has the required spatial and momentum diffusion dependence to result in power law spectra and, upon inserting this $D_0$ into our equation for the power law index $a$, we obtain a spectral index of $-5$. This result has been explored in further detail in \cite{zhang13}.

\subsection{The Inclusion of Convection, Spatial Diffusion and Injection} \label{sec:conv_spa}
This seemingly general result was found in the absence of convection, spatial diffusion, particle injection and charge exchange losses. The inclusion of each of these mechanisms could lead to significant deviation in the spectral index or indeed a change in the spectral shape entirely. As was briefly mentioned in Section \eqref{sec:intro}, ASZ2013 solved the more general equation \eqref{or_tran}, albeit in the absence of losses. Using the diffusion coefficients of equation \eqref{eqn:bal_kappa} in the case with no losses ($\tau_L \rightarrow \infty$), the final form of the steady-state transport equation used in ASZ2013 is given by
\begin{align} \label{parker1}
V_0 \frac{\partial f}{\partial r} = \frac{2 V_0}{3 r} p \frac{\partial f}{\partial p} + \frac{ V_c^2}{5V_0 r^2}\frac{\partial}{\partial r} \left ( r^3 \frac{\partial f}{\partial r} \right )  + \frac{V_0}{3 r} \frac{1}{p^2} \frac{\partial}{\partial p} \left ( p^4 \frac{\partial f}{\partial p} \right ) +Q 
\end{align}
This equation was then solved analytically in ASZ2013 using the scattering time method \cite{schlick02}, to be discussed further in Section \eqref{sec_loss}. For a sensible choice of spatial boundary conditions (an inner reflecting boundary due to the strong magnetic field and an outer free escape boundary due to the weak magnetic field), and if the following conditions are satisfied
\begin{align}
\ln \left(\frac{r_\text{max}}{r_\text{min}}\right) > \frac{2}{5 M_A^2 - 2} && {r_\text{max}} \gg r_\text{min} && M_A > 1
\end{align}
where ${r_\text{max}}$ and $r_\text{min}$ are the maximum and minimum radii respectively and $M_A = V_0/V_c$ is the solar wind Mach number, then this equation retains the $p^{-5}$ solutions at large momenta, with steeper spectra found at lower momenta (see ASZ2013, equation 114 and Figures 1 \& 2 therein).

However, this method must be modified when applied past the termination shock, where adiabatic cooling is considered a negligible affect. Instead, we include the possibility of losses due to charge exchange, a mechanism considered important in the outer heliosphere (see \cite{zhang12}, Figure 9 therein). The pressure growth factor $\gamma$ (and therefore the diffusion coefficients) will be modified with the addition of losses, resulting in the possibility of further spectral changes.  In the next section, we will modify the theory of ASZ2013 by including a loss term. In Sections \eqref{sec_in} and \eqref{sec_down}, we then use this model to determine the resulting spectra in the inner and outer heliosphere respectively.

\section{The Inclusion of Charge Exchange Losses} \label{sec_loss}
We now return to solving the more general transport equation that includes losses, given by equation \eqref{or_tran}. With diffusion coefficients given by equation \eqref{eqn:bal_kappa}, the full transport equation now takes the form
\begin{multline} \label{parker1}
\frac{\partial f}{\partial t} + V_0 \frac{\partial f}{\partial r} = \frac{2 V_0}{3 r} p \frac{\partial f}{\partial p} + \frac{ V_c^2}{5V_0 r^2}\frac{\partial}{\partial r} \left [ \frac{r^3}{1 + 3r/(10V_0\tau_L)} \frac{\partial f}{\partial r} \right ]  \\
+ \frac{V_0(1 + 3r/(10V\tau_L))}{3 r} \frac{1}{p^2} \frac{\partial}{\partial p} \left ( p^4 \frac{\partial f}{\partial p} \right ) +Q -\frac{f}{\tau_L}
\end{multline}
Before attempting to solve equation \eqref{parker1}, we wish to analyse the relevant timescales of each term within it, namely those of convection, adiabatic deceleration, spatial diffusion, momentum diffusion and losses. These are given by
\begin{align}
\tau_C=\frac{r}{V_0} && \tau_A =  \frac{3}{2} \tau_C 
\\\tau_S= 5 M_A^2\left(1 + \frac{3r}{10V_0\tau_L}\right) \tau_C && \tau_M= \frac{3}{1 + 3r/(10V_0\tau_L)} \tau_C && \tau_L
\end{align}
respectively. As we are interested in comparing the timescales of each mechanism, and as it is only the loss term that we have not written in terms of $\tau_C$, we recast it for later convenience as
\begin{equation} \label{eqn:loss_recast}
\tau_L =  \frac{3 \chi}{10(1- \chi)} \tau_C
\end{equation}
or, in terms of $\chi$
\begin{equation} 
\chi = \frac{1}{1+ 3\tau_C/10 \tau_L}
\end{equation}
which is equivalent to implying that $\tau_L \propto r$. With this spatially dependent choice of $\tau_L$, we once again obtain $\kappa = \kappa_0 r$ and $D=D_0 p^2/r$ diffusion coefficients and therefore each term in equation \eqref{parker1} remains in Cauchy-Euler form. Our choice in the scaling factor  has been selected for comparative reasons. In terms of $\chi$, the diffusion coefficients are given by $\kappa = \chi \kappa_\text{ASZ}$ and $D= D_\text{ASZ}/\chi$, where $\kappa_\text{ASZ}$ and $D_\text{ASZ}$ are the diffusion coefficients of equation \eqref{eqn:bal_kappa} in the absence of losses used in ASZ2013. Hence, only the magnitude of the diffusion coefficients are changed in comparison to those of ASZ2013. The quantity $\chi$ is a free parameter which allows us to solve the transport equation for different loss times. In other words, for a loss time that is proportional to $r$, our analysis is different to that of ASZ2013 in two ways: a changing in the magnitudes of the diffusion coefficients, and the inclusion of a loss term.  The timescales now read as 
\begin{align} \label{timescales}
\tau_C=\frac{r}{V_0} && \tau_A =  \frac{3}{2} \tau_C && \tau_S= \frac{5 M_A^2}{\chi} \tau_C && \tau_M= 3 \chi \tau_C && \tau_L =  \frac{3 \chi}{10(1- \chi)} \tau_C
\end{align}

Again, before solving the general transport equation of equation \eqref{parker1}, we wish to solve for an equation similar to that of equation \eqref{eqn:comove} but including losses, namely
\begin{equation} \label{eqn:simple_losses}
 \frac{2 V_0}{3 r} p \frac{\partial f}{\partial p}  + \frac{1}{p^2} \frac{\partial}{\partial p} \left ( p^2 D \frac{\partial f}{\partial p} \right ) - \frac{f}{\tau_L}= 0 
\end{equation} 
Using the momentum diffusion coefficient of equation \eqref{eqn:bal_kappa} and the loss timescale of equation \eqref{eqn:loss_recast}, we once again obtain power law solutions with indices of $-5$ that are \textit{independent of the value of $\chi$}. In other words, no matter what the charge exchange loss rate is, the pressure balance condition adjusts the rate of momentum diffusion in such a way as to retain $p^{-5}$ spectra in all instances.    

However, as in the case with no losses discussed in Section \eqref{sec:conv_spa}, this spectrum will be altered by the inclusion of convection, spatial diffusion and injection. With diffusion coefficients given by equation \eqref{eqn:bal_kappa} and a loss timescale given by equation \eqref{eqn:loss_recast}, the more general steady state transport equation we wish to now solve is given by
\begin{equation} \label{trans1}
V_0 \frac{\partial f}{\partial r} = \frac{2 V_0}{3 r} p \frac{\partial f}{\partial p} + \frac{ V_c^2 \chi}{5V_0 r^2}\frac{\partial}{\partial r} \left ( r^3 \frac{\partial f}{\partial r} \right )  + \frac{V_0}{3 r \chi} \frac{1}{p^2} \frac{\partial}{\partial p} \left ( p^4 \frac{\partial f}{\partial p} \right ) +Q(r,p) -\frac{10 V_0(1-\chi)}{3 \chi r} f
\end{equation}

\subsection{The Scattering Time Method} 

Assuming that the injection term is separable, $Q(r,p) = q_1(r)q_2(p)$, we can rewrite equation \eqref{trans1} as
\begin{equation} \label{te1}
\mathcal L_r f(r,p) +\mathcal  L_p f(r,p) = -\frac{3r \chi q_1(r)}{V_0} q_2(p)
\end{equation}
where
\begin{equation} \label{Lr_1}
\mathcal L_r =\frac{3 \chi^2}{5 M_A^2 r}\frac{d}{dr}r^3\frac{d}{dr} - 3r\chi\frac{d}{dr} - 10(1-\chi)
\end{equation}
is the spatial operator acting on $f$  and 
\begin{equation} \label{Lp_1}
 \mathcal L_p = 2 \chi p \frac{d}{dp} + \frac{1}{p^2}\frac{d}{dp} \left(p^4 \frac{d}{dp} \right)
\end{equation} 
is the momentum operator. Note that the loss term, being independent in both space and momentum, could equally have been placed in the $\mathcal L_p$ operator, with the same results following.  Equation \eqref{te1} can be solved using the ``scattering time'' method. According to this theory, for suitable boundary conditions in space and momentum, this equation can be solved with a solution
\begin{equation} \label{f1}
f(r,p) = \int_0^\infty du \,\,  G(p,u) P(r,u)
\end{equation}
where $G(p,u)$ satisfies 
\begin{equation} \label{eqn:H_bal}
\frac{\partial G}{\partial u} =\mathcal L_p G
\end{equation}
with 
\begin{align} \label{eqn:H_init}
G(p, u= \infty)=0 && G(p,u=0)=q_2(p)
\end{align}
and $P(r,u)$ satisfying
\begin{equation} \label{eqn:M_bal}
\frac{\partial P}{\partial u} =\mathcal L_r P
\end{equation}
with 
\begin{align} \label{init1}
P(r, u= \infty)=0 && P(r,u=0)= \frac{3r \chi q_1(r)}{V_0}
\end{align}
Since $\mathcal L_r$ is of Sturm-Lioville form, $P(r,u)$ can be expanded into an orthonormal system \cite{arf70}
\begin{equation} \label{eqn:M_bal2}
P(r,u) = \sum_i c_i P_i(r) e^{-\lambda_i u}
\end{equation}
where $\lambda_i$ are the eigenvalues of this spatial operator and $c_i$ are the expansion coefficients. Thus, inserting equation \eqref{eqn:M_bal2} into equation \eqref{f1}, we obtain
\begin{equation} \label{f2}
f(r,p) = \sum_i c_i P_i(r) \Pi_i(p)
\end{equation}
where we have defined
\begin{equation} \label{R_def}
\Pi_i(p) \equiv \int_0^\infty du G(p,u) e^{-\lambda_i u}
\end{equation}
Therefore, in order to obtain the particle distribution $f(p,r)$, we need to determine four quantities: the momentum components $\Pi_i(p)$, the spatial components $P_i(r)$, the eigenvalues $\lambda_i$ and the expansion coefficients $c_i$. In Sections \eqref{cal_mom}, \eqref{cal_eigen} and \eqref{cal_exp}, we calculate each of these quantities in turn, before analysing the full solution of equation \eqref{f2} in Section \eqref{anal_results}.

\subsection{Calculating the Momentum Components} \label{cal_mom}
Combining equations \eqref{eqn:H_bal}, \eqref{eqn:M_bal2} and \eqref{R_def} with the initial condition of equation \eqref{eqn:H_init}, we obtain the following ``leaky box equations''
\begin{equation} \label{lp1}
\mathcal L_p \Pi_i(p) - \lambda_i \Pi_i(p) = -q_2(p)
\end{equation}
Inserting $\mathcal L_p$, we recast into the following self-adjoint form 
\begin{equation} \label{app_A1}
\frac{d}{dp} \left(p^{4+2 \chi} \frac{d \Pi_i}{dp} \right)  - \lambda_i p^{2+2\chi} \Pi_i(p) = -  p^{2+2\chi} q_2(p)
\end{equation}
This equation can be solved using Green's functions (see \eqref{app_A}), with solutions
\begin{equation} \label{R1}
\Pi_i(p,p_I) = \frac{Q_0}{2\mu_i p_I }
\begin{cases}
(p/p_I)^{\mu_i - (\chi + 3/2)} & \text{for } p \leq p_I \\
(p/p_I)^{-\mu_i - (\chi + 3/2)} & \text{for } p \geq p_I
\end{cases}
\end{equation}
where we have defined
\begin{equation} \label{mu1}
\mu_i =  \sqrt{\left(\chi + \frac{3}{2} \right)^2 + \lambda_i}
\end{equation}
and we have also assumed that injection is mono-energetic, i.e. $q_2(p_0) = Q_0\delta(p-p_I)$. In the case with no losses, i.e. $\chi = 1$, we obtain
\begin{equation}
\mu_i = \sqrt{\frac{25}{4} + \lambda_i}
\end{equation}
which agrees with equation 44 of ASZ2013. 

\subsection{Calculating the Spatial Components and Eigenvalues} \label{cal_eigen}
Combining equations \eqref{eqn:M_bal2} and \eqref{eqn:M_bal} gives us the following equation for the expansion coefficients $P_i(r)$
\begin{equation}
\mathcal L_r P_i(r) + \lambda_i P_i(r)=0
\end{equation}
Inserting our expression for $\mathcal L_r$ and rearranging, we obtain
\begin{equation} \label{M1}
r^2 \frac{d^2P_i}{dr^2} -2 \eta r \frac{dP_i}{dr} + \Lambda_i P_i(r) = 0
\end{equation}
where we have defined
\begin{align} \label{M1_def}
\eta \equiv \frac{3}{2} \left( \frac{5 M_A^2}{3 \chi} - 1 \right) &&  \Lambda_i \equiv \frac{5 M_A^2 \lambda_i^*}{3 \chi^2} 
\end{align}
and we have shifted the eigenvalues to $\lambda_i^* = \lambda_i - 10(1-\chi)$. This equation is the same as that obtained in ASZ2013 (equation 61 therein), but with redefined expressions for both $\eta$ and $\Lambda_i$. Adopting the same boundary conditions as used in Section \eqref{sum_ant}, and if the following conditions are satisfied
\begin{align}
\ln ({r_\text{max}}/r_\text{min}) > \frac{2}{5 M_A^2 - 2} && {r_\text{max}} \gg r_\text{min} && M_A > 1
\end{align}
we obtain (see \eqref{app_B}) spatial coefficients of the form
\begin{equation} \label{M_cases}
P_i(r) = \begin{cases}
 a_1 r^{\eta+1/2} \sinh[\psi \ln({r_\text{max}}/r)] & i=1\\
 b_1 r^{\eta+1/2} \sin[\nu_i \ln({r_\text{max}}/r)] & i>1
 \end{cases}
\end{equation}
and shifted eigenvalues
\begin{equation} \label{lambda1}
\lambda_i^* = \begin{cases}
\dfrac{3 \chi^2}{5M_A^2}\left(\dfrac{5}{\chi}M_A^2-2\right)^2 \left(\dfrac{r_\text{min}}{{r_\text{max}}}\right)^{5M_A^2/\chi - 2} & i=1\\
\dfrac{3 \chi^2}{5 M_A^2}\left\{(i-1)^2\pi^2 \left[ 1+\dfrac{1}{(5M_A^2/2\chi-1)\ln({r_\text{max}}/r_\text{min})}\right]^2 + \left(\dfrac{5M_A^2}{2\chi} -1\right)^2\right\} & i>1
 \end{cases}
\end{equation}
where we have defined 
\begin{equation} \label{prob_1}
\psi \approx (\eta+1/2)\left[ 1-2\left(\frac{{r_\text{max}}}{r_\text{min}}\right)^{-(1+2\eta)}\right]
\end{equation}
\begin{equation} \label{prob_2}
\nu_i \approx (i-1)\pi \left[ 1+\frac{1}{(\eta+1/2)\ln({r_\text{max}}/r_\text{min})}\right] \,\,\, i=2,3\dotsc
\end{equation} 

\subsection{Calculating the Expansion Coefficients} \label{cal_exp}
Finally, according to equation \eqref{f2}, we need to calculate the expansion coefficients $c_i$ in order to obtain the distribution function (Note that we will absorb the constants $a_1$ and $b_1$ from equation \eqref{M_cases} into these coefficients.). As the $P_i$s form an orthonormal system, they satisfy the orthonormality condition
\begin{equation} \label{M_ortho}
\int_{r_\text{min}}^{r_\text{max}}  r^{-2(\eta+1)} P_m(r) P_n(r) \,\, dr= j_n \delta_{m,n} 
\end{equation}
where
\begin{equation} 
j_i = \int_{r_\text{min}}^{r_\text{max}}  r^{-2(\eta + 1)}  P_i^2(r) \,\, dr
\end{equation}
This relation, coupled with the initial condition given by equation \eqref{init1}, allows us to obtain the following expression for the expansion coefficients
\begin{equation} \label{c1}
c_i = \frac{3 \chi}{V_0 j_i} \int_{r_\text{min}}^{r_\text{max}}  r^{-2\eta - 1} q_1(r) P_i(r)
\end{equation}
For comparative reasons, we also adopt the spatial injection term used in ASZ2013, where they assume pick-up ions are injected in an outer ring distribution of the form
\begin{equation} \label{inj_ant}
q_1(r) = H[r-r_1]H[r_2-r]
\end{equation}
where $r_1=0.5r_{\text{max}}$ and $r_2=0.9r_{\text{max}}$ and $H[n]$ is the Heaviside step function. (This assumption will be relaxed in Sections \eqref{sec_in} and \eqref{sec_down}.) Thus, upon inserting this injection term and the $P_i$s from equation \eqref{M_cases} into equation \eqref{c1} and integrating, we obtain for the expansion coefficients 
\begin{multline} \label{coeff1}
c_1= \frac{3 \chi}{V_0 j_1 {r_\text{max}}^{\eta-\frac{1}{2}}} \left[\psi_1^2 - \left(\eta - \frac{1}{2}\right)^2\right]^{-1} \left \{2^{\eta-\frac{1}{2}}\left[\psi_1 \cosh(\psi_1 \ln 2) \right. \right.\\
\left. \left. - \left(\eta - \frac{1}{2}\right) \sinh (\psi_1 \ln 2)\right]  -\frac{10}{9}^{\eta - \frac{1}{2}} \left[ \psi_1 \cosh \left(\psi_1 \ln \frac{10}{9} \right ) \right. \right. \\
\left. \left. - \left( \eta - \frac{1}{2} \right) \sinh \left ( \psi_1 \ln \frac{10}{9} \right) \right]\right\}
\end{multline}
where
\begin{equation}
j_1 = \frac{\sinh[2 \psi_1 \ln({r_\text{max}}/r_\text{min})]}{4 \psi_1} - \frac{1}{2} \ln({r_\text{max}}/r_\text{min})
\end{equation}
and for $i=2,3 \dotsc$
\begin{multline}  \label{coeff2}
c_i= \frac{3 \chi}{V_0 j_i {r_\text{max}}^{\eta-\frac{1}{2}}} \left[\nu_i^2 + \left(\eta - \frac{1}{2}\right)^2\right]^{-1} \left \{2^{\eta-\frac{1}{2}}\left[\left(\eta - \frac{1}{2}\right) \sin (\nu_i \ln 2) - \nu_i \cos(\nu_i \ln 2) \right] \right. \\
\left. -\frac{10}{9}^{\eta - \frac{1}{2}} \left[ \left( \eta - \frac{1}{2} \right) \sin \left ( \nu_i \ln \frac{10}{9} \right) - \nu_i \cos \left(\nu_i \ln \frac{10}{9} \right ) \right]\right\}
\end{multline}
where
\begin{equation}
j_i = \frac{1}{2} \ln({r_\text{max}}/r_\text{min}) - \frac{\sin[2 \nu_i \ln({r_\text{max}}/r_\text{min})]}{4 \nu_1i} 
\end{equation}

\subsection{Final Distribution Function} \label{anal_results}
Hence, by equation \eqref{f2}, with $\Pi_i(p)$ given by equation \eqref{R1}, $P_i(r)$ given by equation \eqref{M_cases} and $c_i$ given by equations \eqref{coeff1} and \eqref{coeff2}, we obtain the following spectrum
\begin{equation} \label{totes_sol}
f(p,r) = \frac{Q_0}{2p_I} \sum_i \frac{c_i P_i(r)}{\mu_i }
\begin{cases}
(p/p_I)^{\mu_i - (\chi + 3/2)} & \text{for } p \leq p_I \\
(p/p_I)^{-\mu_i - (\chi + 3/2)} & \text{for } p \geq p_I
\end{cases}
\end{equation}

\subsubsection{Analysing $\lambda_1^{*}$}

At high momenta, where the contribution from $\lambda_1$ may dominate over the other eigenvalues if the first expansion coefficient $c_1$ is large enough, we must have that 
\begin{equation} \label{mu2}
\mu_1 = \sqrt{\left(\chi + \frac{3}{2} \right)^2 + \lambda_1^* + 10(1-\chi)} + \chi + \frac{3}{2} =5
\end{equation}
 if we wish to obtain a $p^{-5}$ spectrum. Upon rearranging, this implies that the value for $\lambda_1^*$ must be 
\begin{equation} \label{lambda2}
\lambda_1^* = 0
\end{equation}
Hence, according to equation \eqref{lambda1}, for a particular choice of $\chi$, i.e. for a particular loss rate, the following condition
\begin{equation} \label{cond1}
 \frac{3 \chi^2}{5M_A^2}\left(\frac{5}{\chi}M_A^2-2\right)^2 \left(\frac{r_\text{min}}{{r_\text{max}}}\right)^{5M_A^2/\chi - 2} \ll 1
\end{equation}
must be satisfied to obtain a spectral index of $-5$. ASZ2013 have demonstrated that, for $\chi=1$, this conditions is indeed true. To see if this condition is still true for $\chi \neq 1$, i.e. whether it is still true with the inclusion of losses, we look at three different loss timescales: a long, similar and short timescale in comparison to the convection timescale $\tau_C$. In particular, we calculate equation \eqref{cond1} when the loss timescale $\tau_L$ is equal to $10\tau_C$, $\tau_C$  and $0.1\tau_C$, corresponding to values of $\chi$ equaling $0.97$, $0.77$ and $0.25$ respectively. For each of these loss times, by equation \eqref{lambda1}, we obtain\\
\\
\textbf{Long Timescale} [$\tau_L = 10\tau_C$ ($\chi=0.97$)]: $\lambda_1^{*} =  6.96 \times 10^{-7} \ll 1$\\
\textbf{Similar Timescale} [$\tau_L = \tau_C$ ($\chi=0.77$)]: $\lambda_1^{*} = 2.69 \times 10^{-9} \ll 1$\\
\textbf{Short Timescale} [$\tau_L = 0.1\tau_C$ ($\chi=0.25$)]: $\lambda_1^{*} =  8.66 \times 10^{-34} \ll 1$\\
\\
where we have adopted the same value of $M_A$ ($=1.35$) that was used in ASZ2013. This value of $M_A$ corresponds to very strong turbulence. If we instead choose a larger Mach number, i.e. weaker turbulence, condition \eqref{cond1} becomes even more satisfied. However, the first expansion coefficient $c_1$ becomes smaller, meaning that the momentum at which the spectral index relaxes to $-5$ occurs at a momentum that is much larger than observed. In other words, this choice of Mach number is a best-fit value to match on to the observed spectra. For this choice of Mach number, no matter what the loss timescale is, if we assume $\lambda_1$ dominates over the other $\lambda_i$'s, a $p^{-5}$ spectrum is always achieved at large momenta.

\subsubsection{Analyzing $\lambda_i^*$s, $i=2,3 \ldots$}
\begin{figure}[!h]
\centering
\includegraphics[width= \linewidth]{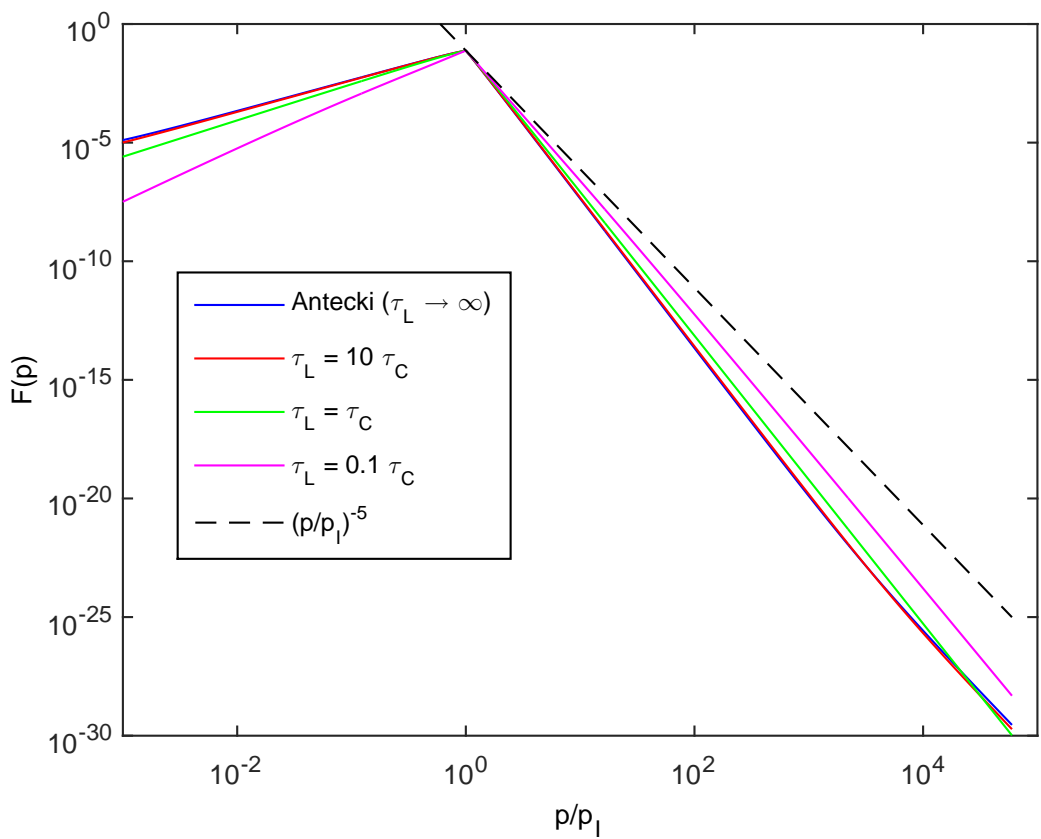}
\caption{The steady state momentum spectra at $r=0.7r_{\text{min}}$ for four different loss times, as determined by equation \eqref{totes_sol}. Each spectra has been normalised according to equation \eqref{totes_norm}. The first 100 eigenvalues and expansion coefficients have been included. Also plotted is a $F(p) \propto p^{-5}$ spectrum for comparison.  }
\label{fig1}
\end{figure}

\begin{figure}[!h]
\centering
\includegraphics[width= \linewidth]{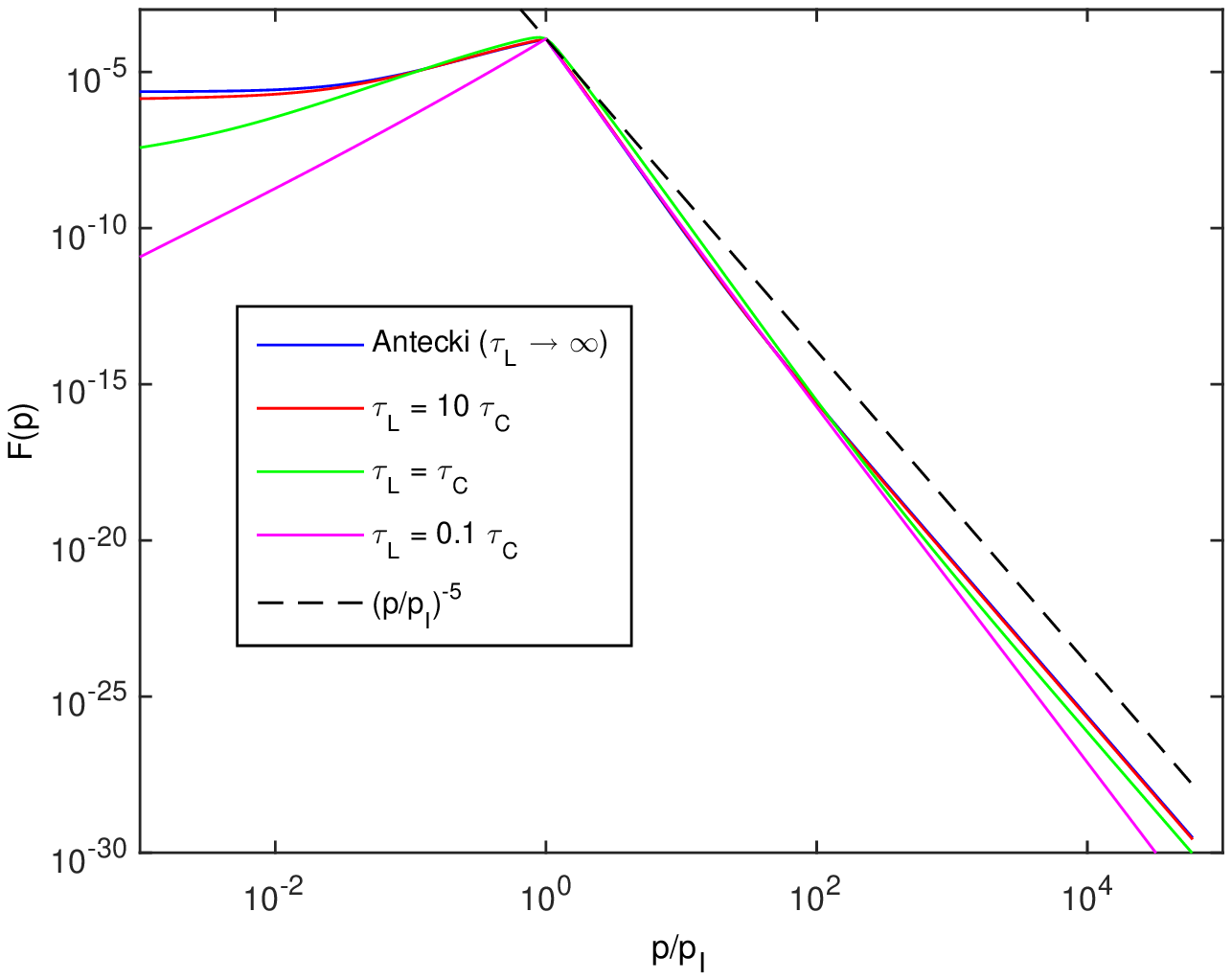}
\caption{The steady state momentum spectra at $r=0.15r_{\text{min}}$ for four different loss times, as determined by equation \eqref{totes_sol}. Each spectra has been normalised according to equation \eqref{totes_norm}. The first 100 eigenvalues and expansion coefficients have been included. Also plotted is a $F(p) \propto p^{-5}$ spectrum for comparison.  }
\label{fig2}
\end{figure}

To see what affect the addition of the other $\lambda_i^*$s has on deviating the spectral index from $-5$ at low momenta, we for clarity list the first $10$ $\lambda_i^*$s, $\mu_i$s and $a_i$s for different loss times in Table \ref{table1A},  where we have recast the expansion coefficients via
\begin{equation} \label{norm_ex}
a_i= \frac{V_0 {r_\text{max}}^{\eta-\frac{1}{2}}}{3} c_i
\end{equation}  
and have defined the spectral power law index as $\mu_i^*= \mu_i + \chi + 3/2$. In Figures \ref{fig1} and \ref{fig2}, we have plotted the the resulting spectra with the inclusion of the first 100 $\lambda_i$'s etc. both inside and outside the source distribution for different loss rates. Note that we have normalized each spectrum to have the same value at the injection momentum in order to easily compare each spectral index to $-5$, i.e. we have normalised each spectrum as
\begin{equation} \label{totes_norm}
F(p)= \frac{2p_I^{3-2\chi} V_0}{3Q_0} \frac{f(\tau_L \rightarrow \infty, p=p_I)}{f(\tau_L, p=p_I)} f(p)
\end{equation}
\begin{table}[!h]
\centering
\begin{tabular}{ |c| c c c| c c c|}
\hline
$ $ &\multicolumn{3}{ |c| }{$\boldsymbol{\tau_L \rightarrow \infty}$} & \multicolumn{3}{ |c| }{$\boldsymbol{\tau_L=10 \tau_C}$} \\
\hline
$\boldsymbol{i}$ & $\boldsymbol{\lambda_i^*}$ & $\boldsymbol{\mu_i^*}$ & $\boldsymbol{a_i}$& $\boldsymbol{\lambda_i^*}$ & $\boldsymbol{\mu_i^*}$ & $\boldsymbol{a_i}$  \\ \hline
$\boldsymbol{1}$ & $1.29\times 10^{-6}$ & $5.00$ & $1.16\times 10^{-5}$ & $6.96\times 10^{-7}$ & $5.00$ & $7.31 \times 10^{-6}$\\ \hline
$\boldsymbol{2}$ & $8.25$ & $6.31$ & $1.27$ & $8.06$ & $6.27$ & $1.33$\\ \hline
$\boldsymbol{3}$ & $20.53$ & $7.67$ & $-0.28$ & $19.53$ & $7.56$ & $-0.30$\\ \hline
$\boldsymbol{4}$ & $40.99$ & $9.37$ & $-0.11$ & $38.66$ & $9.18$ & $-0.13$\\ \hline
$\boldsymbol{5}$ & $69.62$ & $11.21$ & $0.30$& $65.44$ & $10.95$ & $0.33$ \\ \hline
$\boldsymbol{6}$ & $106.45$ & $13.12$ & $-0.31$& $99.87$ & $12.78$ & $-0.32$\\ \hline
$\boldsymbol{7}$ & $151.45$ & $15.06$ & $0.11$ & $141.95$ & $14.65$ & $0.11$ \\ \hline
$\boldsymbol{8}$ & $204.64$ & $17.02$ & $-0.03$ & $191.68$ & $16.54$ & $-0.01$\\ \hline
$\boldsymbol{9}$ & $266.01$ & $19.00$ & $-0.17$& $249.06$ & $18.45$ & $-0.18$ \\ \hline
$\boldsymbol{10}$ & $335.56$ & $20.99$ & $0.13$  & $314.09$ & $20.37$ & $0.14$\\ \hline
\multicolumn{7}{ |c| }{}\\ \hline
$ $ & \multicolumn{3}{ |c| }{$\boldsymbol{\tau_L=\tau_C}$}  & \multicolumn{3}{ |c| }{$\boldsymbol{\tau_L=0.1 \tau_C}$} \\ \hline
$\boldsymbol{i}$ & $\boldsymbol{\lambda_i^*}$ & $\boldsymbol{\mu_i^*}$ & $\boldsymbol{a_i}$& $\boldsymbol{\lambda_i^*}$ & $\boldsymbol{\mu_i^*}$ & $\boldsymbol{a_i}$  \\ \hline
$\boldsymbol{1}$  & $2.69\times 10^{-9}$ & $5.00$ & $1.11 \times 10^{-7}$ & $8.66\times 10^{-34}$ & $5.00$ & $0.02$\\ \hline
$\boldsymbol{2}$ & $7.00$ & $6.07$ & $2.00$ & $6.32$ & $5.86$ & $947.47$ \\ \hline
$\boldsymbol{3}$  & $13.83$ & $6.88$ & $-0.63$ & $6.96$ & $5.94$ & $-747.71$ \\ \hline
$\boldsymbol{4}$   & $25.21$ & $7.99$ & $-0.33$ & $8.03$ & $6.06$ & $-109.16$\\ \hline
$\boldsymbol{5}$   & $41.15$ & $9.24$ & $0.66$ & $9.52$ & $6.23$ & $742.30$\\ \hline
$\boldsymbol{6}$   & $61.64$ & $10.58$ & $-0.53$ & $11.44$ & $6.44$ & $-586.93$\\ \hline
$\boldsymbol{7}$ & $86.69$ & $11.97$ & $0.10$ & $13.79$ & $6.68$ & $-29.19$\\ \hline
$\boldsymbol{8}$  & $116.29$ & $13.39$ & $0.17$ & $16.56$ & $6.96$ & $511.79$\\ \hline
$\boldsymbol{9}$   & $150.44$ & $14.83$ & $-0.39$ & $19.77$ & $7.26$ & $-494.34$\\ \hline
$\boldsymbol{10}$ & $189.14$ & $16.29$ & $0.25$ & $23.39$ & $7.58$ & $91.44$\\ \hline
\end{tabular}
\caption{Spatial eigenvalues $\lambda_i$, spectral indices $\mu_i^*$ and normalised expansion coefficients $a_i$, as defined in equation \eqref{norm_ex}, for different loss times, where we have assumed $M_A = 1.35$ and ${r_\text{max}}=10r_\text{min}$}. 
\label{table1A}
\end{table}
\paragraph{Inside the injection zone (Figure \ref{fig1}):}
Above the injection momentum, all spectral indices are harder than $-5$ at low momenta. However, with increasing loss rate, the spectra are softer, resulting in spectra closer to $p^{-5}$. At larger momenta, the contribution from $i >1$ eigenvalues become less important and the spectra soften back towards a $-5$ index, as is evident with the blue and red spectra. However, for even larger loss times, this softening is not observed. This can be be explained by comparing the timescales for both momentum diffusion and losses given by equation \ref{timescales}, namely
\begin{align}
\tau_M = 3 \chi \tau_C && \tau_L = \frac{3 \chi}{10(1- \chi)} \tau_C
\end{align}
The loss mechanism becomes faster than momentum diffusion ($\tau_L < \tau_M$) for values of $\chi$ that satisfy $\chi > 0.9$. Hence, according to equation \ref{timescales}, this corresponds to loss times $\tau_L < 3.9 \tau_C$. In this limit, which both the green and pink spectra satisfy, losses dominate over momentum diffusion and softening at high momenta does not occur.

\begin{figure}[!h]
  \begin{subfigure}[b]{0.5\linewidth}
    \centering
    \includegraphics[width=\linewidth]{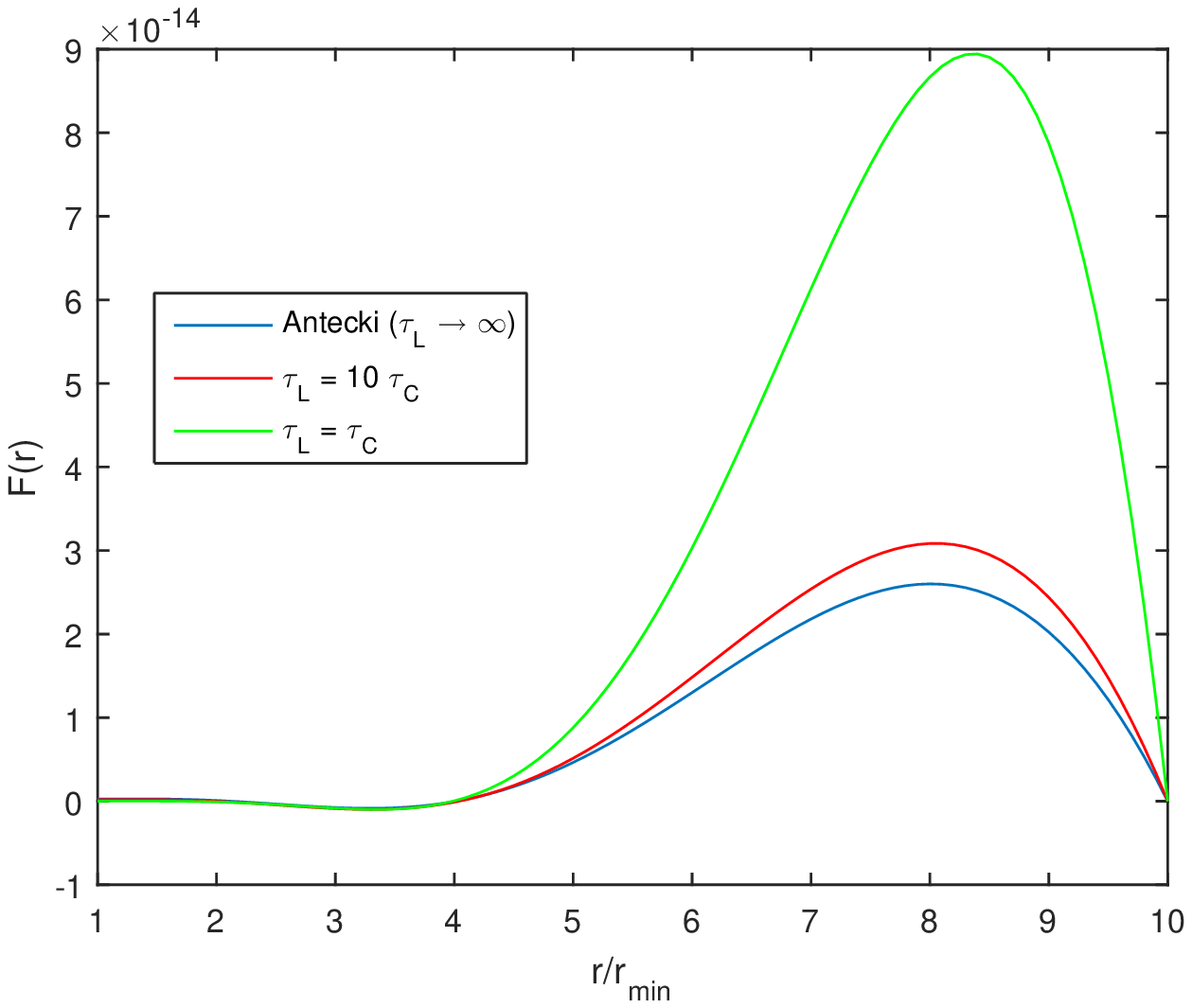} 
    \label{large1} 
  \end{subfigure}
  \begin{subfigure}[b]{0.5\linewidth}
    \centering
    \includegraphics[width=\linewidth]{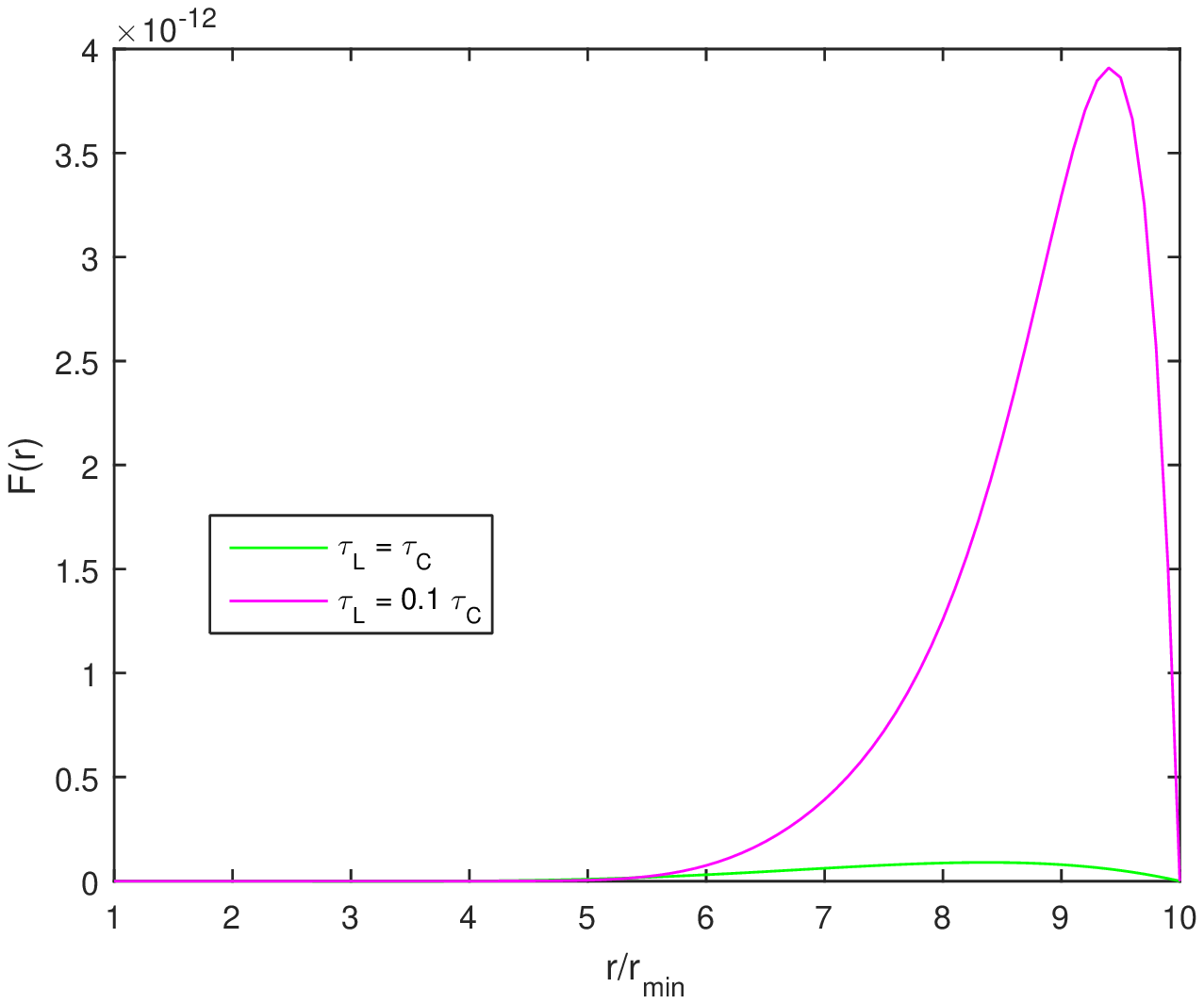} 
    \label{large2} 
  \end{subfigure} 
  \caption{The steady state radial profiles at $p=100 p_I$ for four different loss times, as determined by equation \eqref{totes_sol}. We have separated the $\tau_L = 0.1 \tau_C$ case as the amplitude is  much larger than the other profiles and plotting it in the same figure would suppress the features of the other profiles. Each spectra has been normalised according to equation \eqref{totes_norm2}. The first 100 eigenvalues and expansion coefficients have been included. The Gibbs phenomenon is observed at $r/r_\text{min} \approx 2-4$.  }
  \label{sub1} 
\end{figure}

\begin{figure}[!h]
\centering
\includegraphics[width= \linewidth]{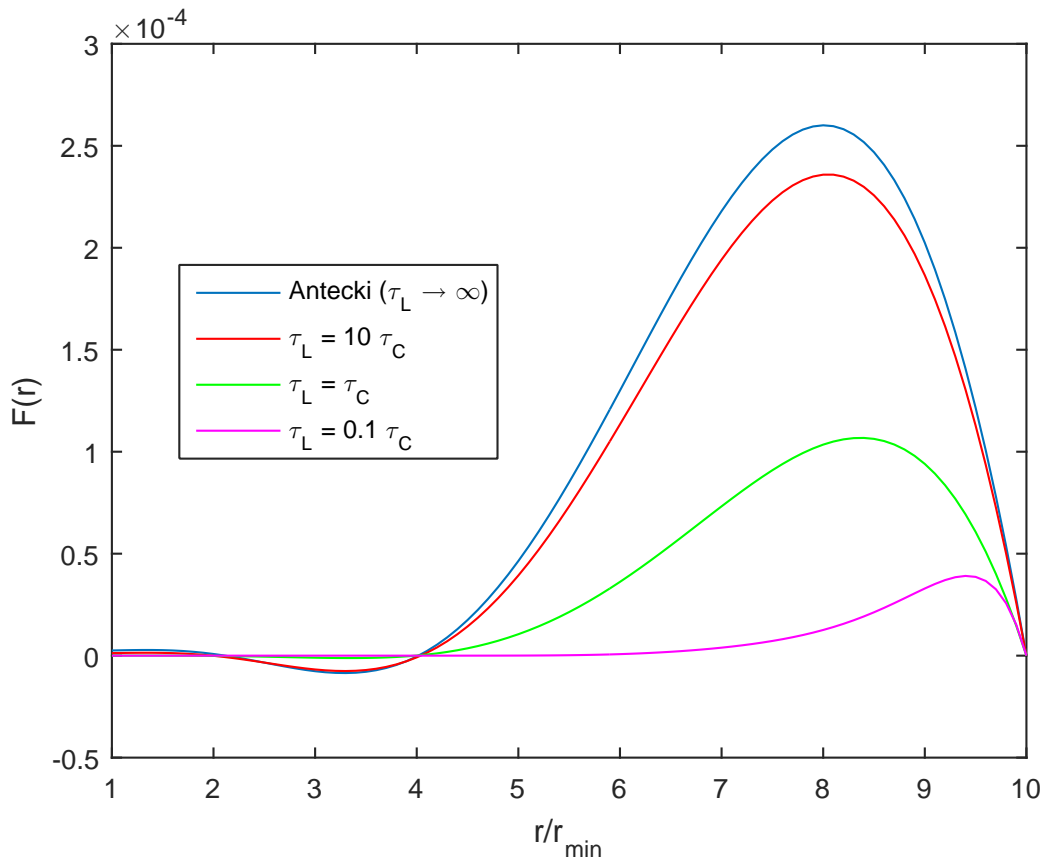}
\caption{The steady state radial profiles at $p=0.01 p_I$ for four different loss times, as determined by equation \eqref{totes_sol}. Each spectra has been normalised according to equation \eqref{totes_norm2}. The first 100 eigenvalues and expansion coefficients have been included. The Gibbs phenomenon is observed at $r/r_\text{min} \approx 2-4$.}
\label{sub2}
\end{figure}

\paragraph{Outside the injection zone (Figure \ref{fig2}):}
Below the injection momentum, we find a more complicated spectra than was evident inside the injection zone. However, with an increasing loss rate (i.e. as losses begin to dominate), a return to a power law shape is found. Above the injection momentum, we once again find spectra that are softer than $-5$. At low momenta, the green curve, corresponding to a loss time equal to that of the convective timescale, has the softest slope. It would appear that the spectral indices soften towards $-5$ at higher momenta, as is evident by the blue, red and green spectra. However, at an even greater loss rate, as with the pink spectrum, this softening does not occur as once again losses occur at a faster rate compared to momentum diffusion. The softening of the spectra towards $-5$ also appears to occur at a lower momentum than spectra found inside the injection zone.
\vspace{10mm}\\
In Figures \ref{sub1} and \ref{sub2}, we have plotted the radial profiles at momenta both above and below the injection momentum, where we have once again included the first 100 $\lambda_i$'s. For these spatial plots, we have normalized the spectra as
\begin{equation} \label{totes_norm2}
F(r)= \frac{2p_I^{3-2\chi} V_0}{3Q_0} f(r)
\end{equation}  

\paragraph{Above the injection momentum (Figure \ref{sub1}):}
In all four cases, as we would expect, most particles are found at large radii, both due to the placement of the injection zone and due to the reflecting boundary at the minimum radius. With an increasing loss rate, the intensity of particles increases. This is perhaps a counter-intuitive result, as we would expect there to be less particles with energies above the injection momenta if there are more losses. However, the loss time also changes the magnitude of the momentum diffusion coefficient due to our pressure balance condition, and thus can enhance particle acceleration. In other words, if there are particles (and therefore energy) lost from the system, this can be balanced by the remaining particles being further accelerated. Also, with increasing losses, the maximum intensity appears to be increasing to higher radii.

At small radii, the distribution appears to be negative, which is of course not possible. This abnormality is also found in ASZ2013 (Section 6 therein) and is due to the Gibbs phenomenon. According to this theory, the eigenfunction series of a sharp discontinuity can both undershoot or overshoot, creating this artifact. Our choice of spatial injection term of equation \eqref{inj_ant} falls under this category, resulting in the observed undershooting at small radii.

\paragraph{Below the injection momentum (Figures \ref{sub2}):} 
In these cases, the opposite affect appears to be occurring. As we increase losses, i.e. remove energy from the system, pressure balance can be sustained if particles are accelerated to energies above the injection momentum. This, in turn, will lead to less particles and therefore lower particle intensities below the injection momentum. However, as was the case above $p_I$, the maximum intensity once again shifts to larger spatial distances with increasing losses.

\vspace{10mm}
We have shown that the inclusion of charge exchange losses can alter the simplified $p^{-5}$ spectrum further from that which is obtained in the absence of convection, spatial diffusion and injection. However, to achieve these results analytically, numerous assumptions were required.  In the next chapter, we take a numerical approach, allowing us to remove some of these assumptions and in turn apply our results to the heliosphere.

\section{Application to the Inner Heliosphere} \label{sec_in}

The analytical work of Section \eqref{sec_loss} has allowed us to demonstrate that a $p^{-5}$ spectrum, as well as deviations from it, are indeed possible under this pressure balance condition. However, in order to analytically solve the transport equation, as given in equation \eqref{parker1}, a number of key assumptions were made
\begin{itemize}
\item A constant solar wind speed ($\mathbf{V} = V_0 \hat{r}$) throughout the acceleration region
\item The spatial component of the injection term, as stated in equation \eqref{inj_ant}, takes the form $q_1(r) = H[r-r_1] H[r-r_2]$
\item A spatially dependent loss time of the form $\tau_L(r) \propto r$
\end{itemize}
A constant solar wind speed is considered a good approximation for the inner heliosphere (in Section \eqref{sec_down}, where we look beyond the termination shock, this assumption will be replaced by a $V \propto 1/r^2$ approximation). However, the final two assumptions shall be replaced by more accurate approximations for the inner heliosphere to see what affect, if any, it has on the steady state spectra. Other assumptions, e.g. spherical symmetry, mono-energetic injection and of course the validity of the quasi-linear approach, we still assume to be valid. 

The form of the spatial injection term used in Section \eqref{sec_loss}, namely that of equation \eqref{inj_ant}, was an approximation chosen so as to easily compare our results to those of ASZ2013. In this section, we relax this restriction and instead use a the more accurate spatial injection term for pick-up ions as given in \cite{chalov04} (equation 10 therein), namely
\begin{equation} \label{inj_acc}
q_1(r) = \frac{\beta_{iE} n_{H \infty}}{r^2} \exp \left(- \frac{\beta_{i E} \text{AU}^2}{r V_{\text{ISM}}} \right)
\end{equation}
where $\beta_{iE}$ is the ionisation rate of hydrogen at $1$ AU, $n_{H \infty}$ is the hydrogen density at the outer radius and $V_{\text{ISM}}$ is the speed of hydrogen relative to the Sun.

Assuming losses are due to charge exchange, the spatial variation of the loss timescale is well understood (see \cite{zhang12}, Figure 9 therein). A timescale of the form $\tau_L \propto r$ is well approximated for small heliospheric distances. At even smaller distances, losses by charge exchange are considered negligible. At large distances, including past the termination shock, Figure 9 of \cite{zhang12} infers that a constant loss time would be a  more accurate approximation. Therefore, a good approximation for the loss time by charge exchange in the inner heliosphere is given by
\begin{equation}
\tau_L(r) = 
\begin{cases}
\infty & 0.01 \text{ AU} < r < 5 \text{ AU} \\
10^{3} \left(\dfrac{r}{10 \text{ AU}}\right) \tau_{C0} &  5 \text{ AU}< r <10 \text{ AU}  \\
10^{3} \tau_{C0} &  10 \text{ AU}< r < 85 \text{ AU}
\end{cases}
\end{equation}
where $\tau_{C0} = 1 \text{ AU}/V_0$ and the large loss time of $\tau_L = 10^3 \tau_{C0}$ corresponds to $\chi \approx 0.9997$.
We can combine these three types into one form, given by
\begin{equation} \label{tau_compact}
\tau_L = \frac{3 \chi}{10(1-\chi)} \left(\frac{r}{10 \text{ AU}} \right)^\sigma \frac{1 \text{ AU}}{V_0} \,\,\,\,\,\, \sigma \in \{0,1\}, \,\,\, \chi \in \{0.9997,1\}
\end{equation}
where $\chi =1$ refers to the first spatial range with no losses, $\chi = 0.9997$ \& $\sigma = 1$ represents the second range where $\tau_L \propto r$, and $\chi = 0.9997$ \& $\sigma = 0$ corresponds to the third range of a constant loss time, where we have once again chosen the proportionality constant so as to easily compare to the work of ASZ2013. 

This more general loss time results in a new form of spatial diffusion coefficient given by
\begin{equation} \label{kappa_h}
\kappa(r) = \frac{V_c^2 r \chi}{5 V_0} h(r)\,\,\,\,\,\, \sigma \in \{0,1\}, \,\,\, \chi \in \{0.9997,1\}
\end{equation}
where 
\begin{equation} \label{h_par}
h(r) =\left\{ \chi + 10(1-\chi) \left(\frac{r}{10 \text{ AU}} \right) ^{-\sigma+1} \right\} ^{-1}
\end{equation}
Once again note that when $\chi = 1$, this form of $\kappa$ reduces to that of equation \eqref{eqn:bal_kappa}. The corresponding forms of both the spatial operator, as previously given by equation \eqref{Lr_1}, and the momentum operator, as previously given by equation \eqref{Lp_1}, are
\begin{equation} 
\mathcal L_r =\frac{3 \chi^2 h(r)}{5 M_A^2 r}\frac{d}{dr}\left(r^3 h(r)\frac{d}{dr}\right) - 3r\chi h(r) \frac{d}{dr} - 100h(r)(1-\chi) \left(\frac{r}{10 \text{ AU}} \right) ^{-\sigma+1}
\end{equation}
\begin{equation}
\mathcal L_p =2 \chi h(r) p \frac{d}{dp} + \frac{1}{p^2}\frac{d}{dp} \left(p^4 \frac{d}{dp} \right)
\end{equation}
However, note that the momentum operator is now, in general, no longer spatially independent. Therefore, the scattering time method introduced in Section \eqref{sec_loss} can no longer be applied. Instead, we solve the transport equation, given by equation \eqref{parker1}, with spatial and momentum diffusion coefficients given by equations \eqref{kappa_h} and \eqref{eqn:bal_D} respectively, numerically using the Gauss Seidel finite difference method (see \eqref{app_C} for further details).

\begin{figure}[!h]
\centering
\includegraphics[width= \linewidth]{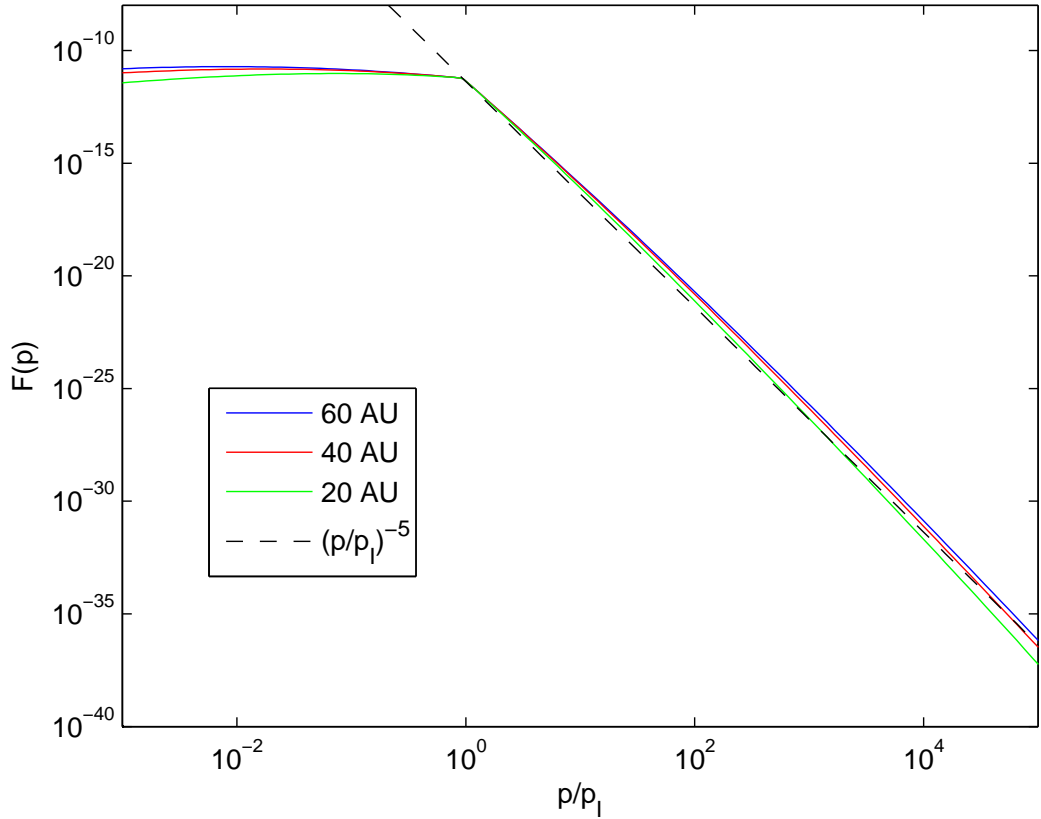}
\caption{The steady state momentum spectra at three different spatial distances within the inner heliosphere for the choice of parameters described in Section \eqref{sec_in}, where we have adopted $M_A=1.35$.  Each spectra has been normalised to the case of $\tau_L \rightarrow \infty$ in order to better compare the spectral indices. Also plotted is a $F(p) \propto p^{-5}$ spectrum for comparison. }
\label{inner_midM}
\end{figure}

\begin{figure}[!h]
\centering
\includegraphics[width= \linewidth]{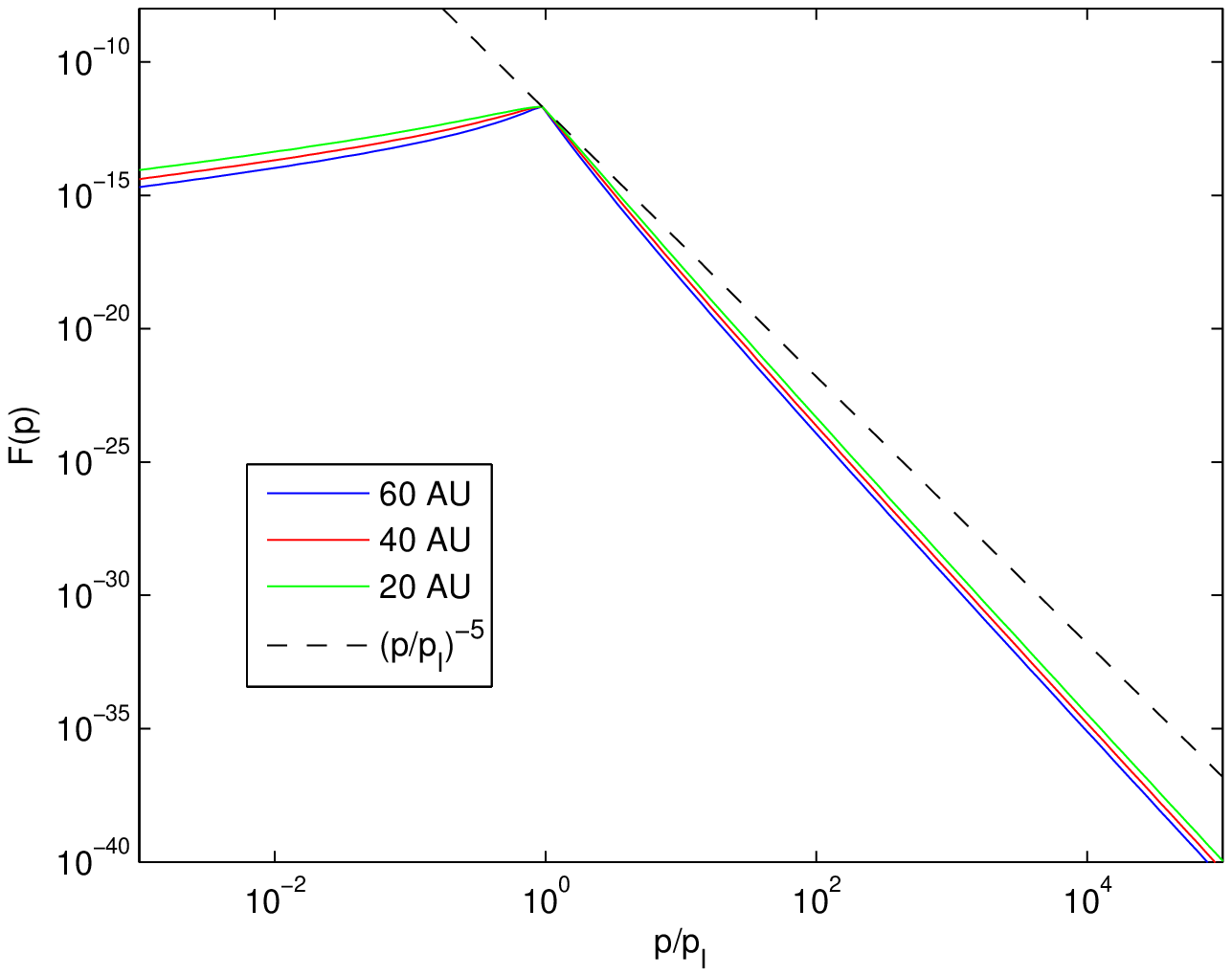}
\caption{The steady state momentum spectra at three different spatial distances within the inner heliosphere for the choice of parameters described in Section \eqref{sec_in}, where we have adopted $M_A=0.35$.  Each spectra has been normalised to the case of $\tau_L \rightarrow \infty$ in order to better compare the spectral indices. Also plotted is a $F(p) \propto p^{-5}$ spectrum for comparison. }
\label{inner_smallM}
\end{figure}

Figure \ref{inner_midM} presents the resulting spectra at three different positions for a Mach number $M_A=1.35$. As the loss timescale is very long compared to other competing process, we expect its inclusion to have little to no affect in changing the spectra from those found in ASZ2013. However, the spectra of Figure \ref{inner_midM} appear to differ to those found in ASZ2013 and Section \eqref{sec_loss}; instead, the more complicated spectral structure previously found is suppressed in favour of a more universal $p^{-5}$ spectra shape above the injection momentum. This is caused primarily by to the choice of the more accurate spatial injection term. Deviations from indices of $-5$ can only be obtained for unlikely very small values of $M_A$, corresponding to very strong turbulence - see Figure \ref{inner_smallM} where we have repeated the process for $M_A = 0.35$.

\section{Beyond the Termination Shock} \label{sec_down}
\begin{figure}[!h]
\centering
\includegraphics[width= \linewidth]{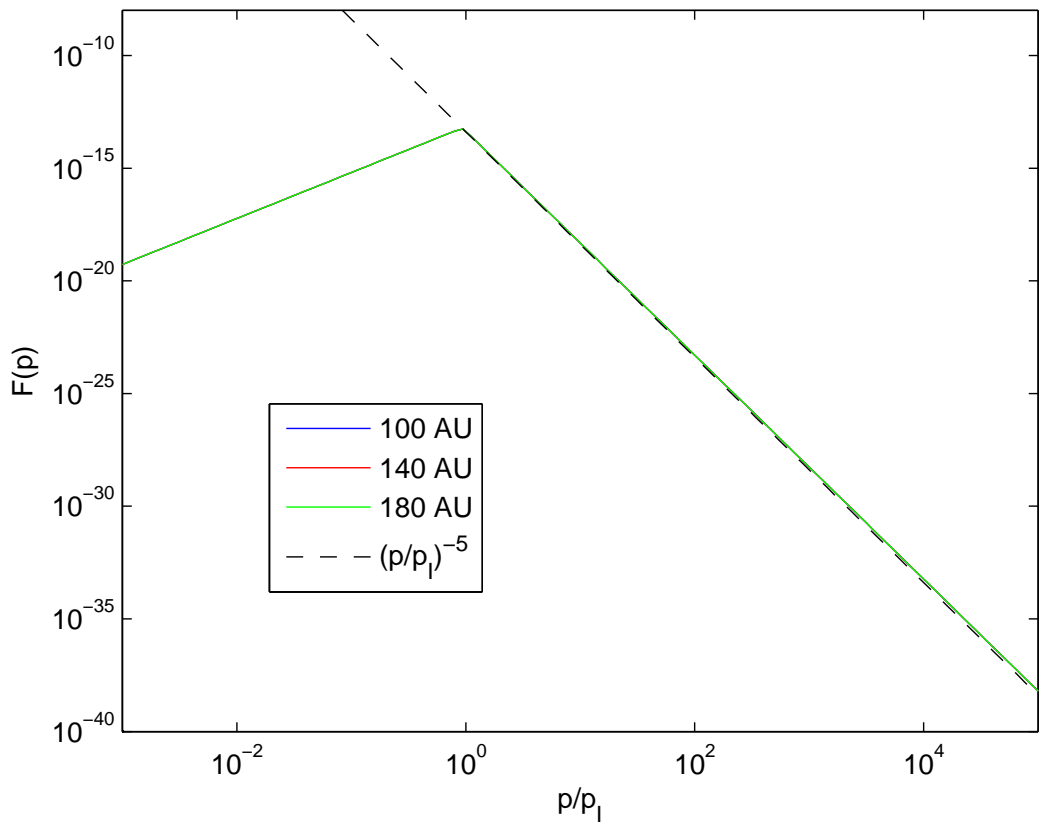}
\caption{The steady state momentum spectra at three different spatial distances beyond the termination shock for the choice of parameters described in Section \ref{term_sh}, where we have adopted $M_A=1.35$. Note that all three spectra overlap over the entire range. Each spectra has been normalised to the case of $\tau_L \rightarrow \infty$ in order to better compare the spectral indices. Also plotted is a $F(p) \propto p^{-5}$ spectrum for comparison. }
\label{outer_midM2}
\end{figure}

In this region, where adiabatic decleration is considered neglible, we adopt the sensible velocity profile \cite{zhang12}
\begin{align}
\mathbf{V}(r) =  \frac{V_0}{R} \left(\frac{85 \text{ AU}}{r}\right)^2 \mathbf{\hat{r}} \,\,\,\,\,\,\, 85 \text{ AU}<r<200 \text{ AU}
\end{align}
where $R$ is the compression ratio of the termination shock ($\approx 2$), whose location we have taken to be at $85 \text{ AU}$. The injection rate in this region is taken to be $R \,q_1({r=85 \text{ AU}})$ \cite{zhang12}. A sensible loss time of the form
\begin{align}
\tau_L(r) = 10^3 \tau_{C0}\,\,\,\,\,\,\, 85 \text{ AU}<r<200 \text{ AU}
\end{align}
is chosen, which again implies a very small loss rate. However, as we have assumed cooling is unimportant in the heliosphere, momentum diffusion is now balanced only by losses. Adopting a Mach number of $M_A = 1.35$, the values of $h(r)$ defined in equation \ref{h_par} varies from $0.005 - 0.0118$. Therefore, there is a large reduction in the spatial diffusion coefficient which in turn, according to equation \ref{eqn:bal_D}, leads to a large increase in the momentum diffusion coefficient. As the momentum diffusion process now dominates, $p^{-5}$ spectra are still easily attained - see Figure \ref{outer_midM2}.

\section{Conclusions}
In order to explain the apparent universal $p^{-5}$ tails  observed through the heliosphere, we have appealed to stochastic acceleration as the possible explanation of their existence. Many different forms of stochastic acceleration exist, depending on the type of turbulence involved.  However, \cite{zhang13} have demonstrated that if the turbulence is composed of small scale magnetohydrodynamic waves, stochastic acceleration is not fast enough to overcome the affect of adiabatic cooling. Instead, we have appealed to large-scale modes; in particular, fluctuations of a compressible nature. This form of acceleration is maximised when both the spatial and momentum diffusion coefficients are related via equation \eqref{eqn:bal_kappa}. Adopting a so called ``pressure balance'' concept between momentum diffusion and adiabatic deceleration to obtain these coefficients, we found that power law spectra with $-5$ spectra indices naturally arise throughout the heliosphere in environments far from sources and where convection and spatial diffusion are considered negligible. 

As was seen in Section \eqref{sec_loss}, the inclusion of both these mechanisms and charge exchange losses can lead to a steepening of the spectra, particularly at low momenta. However, for sensible choices of the free parameters, boundary conditions and injection term, we found in Section \eqref{sec_in} that these features are suppressed. Except for the unlikely case of very strong turbulence, $p^{-5}$ spectra are still obtained above the injection momentum.

If we only consider pressure balance between both momentum diffusion and adiabatic cooling, it alone cannot explain the presence of the suprathermal tails in the outer heliosphere, where cooling is negligible. Instead, in Section \eqref{sec_down}, we considered a balance between momentum diffusion and charge exchange losses. Again, for realistic values of the the solar wind Mach number, $p^{-5}$ spectra are obtained in the outer heliosphere, independent of both the loss rate and distance form the termination shock. 

However, in order to obtain the spatial diffusion coefficient of equation \eqref{eqn:bal_kappa}, an unlikely momentum independent spatial diffusion coefficient was assumed. Dropping this assumption results in a complicated integro-differential equation for the particle pressure. It would be interesting to see if a workaround could be found to obtain a spatial diffusion coefficient that is both momentum and spatially dependent using this notion of pressure balance.

We have also approximated the more exact spatial dependence of the loss time as found in \cite{zhang12}, Figure 9 therein. However, if we assume that losses are by charge exchange, then this loss time is also energy (and therefore momentum) dependent (see \cite{zhang12}, Figure 2 therein). Once again, this leads to similar problems in adopting the pressure balance notion as is found with a momentum dependent spatial diffusion coefficient.

Also, again according to Figure 9 of \cite{zhang12}, the Mach number $M_A$ is not spatially independent as we assumed; rather, it varies throughout the heliosphere. However, as we discovered in Section \eqref{sec_in}, the resulting spectra are not sensitive to this choice of $M_A$ except in unlikely cases of very small values corresponding to very strong fluctuations. We therefore do no believe the inclusion of a spatially dependent Mach number will have much affect on our results.

One particular feature of the suprathermal tail that cannot be explained by our theory is that of the observed step feature (see \cite{fahr12} - Figure 1 therein). This sharp drop at the injection momentum has not been obtained in any of our analyses. However, the bimodal treatment used in \cite{zhang12} naturally lead to the creation of this feature. It would also be interesting to see if a bimodal approach to our work could also lead to the creation of this step feature.

Finally, we have applied this notion of pressure balance to only one particular branch of turbulence, namely large-scale compressions, in only one particular setting, namely the heliosphere. An application of this notion to explain other unresolved cosmic ray phenomena, both within the heliosphere and indeed elsewhere,  may lead to interesting insights.
\vspace{10mm}\\
C.K. and P.D. acknowledges support of this work by the IRC, formerly IRCSET, through grant R11673.

\appendix

\section{Solving Equation \eqref{app_A1} using Green's Functions} \label{app_A}
We begin with the equations we wish to solve, namely equation \eqref{app_A1}
\begin{equation}
\frac{d}{dp} \left(p^{4+2 \chi} \frac{d \Pi_i}{dp} \right)  - \lambda_i p^{2+2\chi} \Pi_i(p) = -  p^{2+2\chi} q_2(p)
\end{equation}
These equations can be solved using Green's functions
\begin{equation} \label{R10}
\Pi_i(p) = \int_0^\infty dp_0  p_0^{2+2\chi} q_2(p_0) \Gamma_i(p,p_0)
\end{equation}
where $\Gamma_i(p,p_0)$ satisfies 
\begin{equation} \label{G1}
\frac{d}{dp} \left(p^{4+2\chi} \frac{d \Gamma_i}{dp} \right)  - \lambda_i p^{2+2\chi} \Gamma_i = -  \delta(p-p_0)
\end{equation}
We trial power law solutions to the above equations, namely that $\Gamma_i(p,p_0) = A_i(p_0)p^{a_i}$. Inserting this into the above, we obtain $a_i^2+5a_i-\lambda=0$ as the equations for the $a_i$'s. These equations have solutions $a_i = -(\chi + 3/2) \pm \mu_i$ where $\mu_i$ depends on both $\lambda_i$ and $\chi$ via
\begin{equation} \label{mu1}
\mu_i =  \sqrt{\left(\chi + \frac{3}{2} \right)^2 + \lambda_i}
\end{equation}
Thus our Green's functions solution are currently
\begin{equation}
\Gamma_i(p,p_0) =
\begin{cases}
A_i(p_0)p^{-(\chi+3/2)+\mu_i} + B_i(p_0)p^{-(\chi+3/2)-\mu_i} & p \leq p_0 \\
C_i(p_0)p^{-(\chi+3/2)+\mu_i} + D_i(p_0)p^{-(\chi+3/2)-\mu_i} & p \geq p_0
\end{cases}
\end{equation}
If we use the following sensible momentum boundary conditions
\begin{align}
f(r,p=0) = \text{finite} && f(r, p \rightarrow \infty) = 0
\end{align}
i.e. that there are a finite number of particles with no energy and no particles with infinite energy, then this implies that $B_i(p_0) = C_i(p_0)=0$. Thus our solutions are reduced to
\begin{equation}
\Gamma_i(p,p_0) =
\begin{cases}
A_i(p_0)p^{-(\chi+3/2)+\mu_i}  & p \leq p_0 \\
D_i(p_0)p^{-(\chi+3/2)-\mu_i} & p \geq p_0
\end{cases}
\end{equation}
We must also have continuity at $p=p_o$, implying that $A_i(p_0)p_0^{-(\chi+3/2)+\mu_i}=D_i(p_0)p_0^{-(\chi+3/2)-\mu_i}$, i.e. that $A_i(p_0) = D_i(p_0) p_0^{-2\mu_i}$. We also have a jump condition at $p=p_0$ due to the singular behaviour at the discontinuity.  Upon integration, this condition implies that
\begin{multline}
p_0^{4+2\chi}[(-(\chi+3/2)-\mu_i)D_i(p_0)p^{-(\chi+3/2)-\mu_i -1} \\
- (-(\chi+3/2)+\mu_i)A_i(p_0)p^{-(\chi+3/2)+\mu_i -1}] = -1
\end{multline}
Inserting that $A_i(p_0) = D_i(p_0) p_0^{-2\mu_i}$ and rearranging, we obtain for both $A_i(p_0)$ and $D_i(p_0)$
\begin{align}
A_i(p_0) = \frac{1}{2\mu_i}p_0^{-(\chi+3/2)-\mu_i} && D_i(p_0) = \frac{1}{2\mu_i}p_0^{-(\chi+3/2)+\mu_i}
\end{align}
Hence, the solution to equation \eqref{G1}  is given by
\begin{equation}
\Gamma_i(p,p_0) = \frac{(p p_0)^{-(\chi +3/2)}}{2\mu_i} 
\begin{cases}
(p/p_0)^{\mu_i} & \text{for } p \leq p_0 \\
(p/p_0)^{-\mu_i} & \text{for } p \geq p_0
\end{cases}
\end{equation}
Inserting this expression for $\Gamma_i(p,p_0)$ back into equation \eqref{R10}, we obtain
\begin{equation}
\Pi_i(p) = \frac{1}{2 \mu_i p^{\chi + 3/2}} \left[p^{- \mu_i} \int_0^p dp_0  p_0^{\chi + 1/2 + \mu_i} q_2(p_0) + p^{\mu_i} \int_p^\infty dp_0  p_0^{\chi + 1/2 - \mu_i} q_2(p_0) \right]
\end{equation}
If we assume that injection is mono-energetic, namely that $q_2(p_0) = Q_0\delta(p-p_I)$, we obtain the following solution for $\Pi_i$
\begin{equation}
\Pi_i(p,p_I) = \frac{Q_0}{2\mu_i p_I }
\begin{cases}
(p/p_I)^{\mu_i - (\chi + 3/2)} & \text{for } p \leq p_I \\
(p/p_I)^{-\mu_i - (\chi + 3/2)} & \text{for } p \geq p_I
\end{cases}
\end{equation}
as required. 

\section{Solutions to Equation \eqref{M1}} \label{app_B}
We begin with the equation we wish to solve, namely equation \eqref{M1}
\begin{equation} \label{M1_app}
r^2 \frac{d^2P_i}{dr^2} -2 \eta r \frac{dP_i}{dr} + \Lambda_i P_i(r) = 0
\end{equation}
where we have defined both $\eta$ and $\Lambda_i$ in equation \eqref{M1_def}. Recasting equation \eqref{M1_app} with $P_i(r) = r^{\eta} \rho(r)$, we obtain
\begin{equation} \label{eqn:Es}
r^2 \frac{d^2 \rho}{dr^2} +\frac{1}{4} \rho + \left[\Lambda_i - \left( \eta + \frac{1}{2}\right) ^2 \right] \rho = 0
\end{equation}
This equation is solvable as $\rho(r) \propto r^k$ where $k$ satisfies
\begin{equation}
\left( k - \frac{1}{2} \right)^2 = \left( \eta + \frac{1}{2} \right)^2 - \Lambda_i
\end{equation}

\subsection*{Case 1: $\Lambda_i < (\eta+ 1/2)^2$}
We are primarily interested in the smallest $\lambda_i$'s (which in turn is when $\Lambda_i < (\eta + 1/2)^2$) as these eigenvalues will dominate the spectrum at high momenta. Setting $\psi^2 = ( \eta + 1/2 )^2 - \Lambda_i >0$,  the general solution to equation \eqref{eqn:Es} is then
\begin{equation}
\rho(r) = r^{1/2}(a_1 r^\psi + a_2 r^{- \psi})
\end{equation}
To find $a_1$ and $a_2$, suitable spatial boundary conditions need to be chosen. We adopt the same spatial range as is used in ASZ2013, namely a minimum value of $r_0$ and a corresponding maximum value of $10 r_0$. According to the Parker spiral model of the solar magnetic field, a $B(r) \propto 1/r^2$ spatial dependence is a suitable approximation for the inner heliosphere. Hence, at the inner boundary where the magnetic field is strong, a reflecting boundary of the form $(dP/dr)_{r_0} =0$ is a sensible choice. At the outer boundary, where the magnetic field is much weaker, particles can more easily escape the region, and therefore a free escape boundary ($P(R)=0$) is chosen. The second condition implies that 
\begin{equation}
a_2 = -a_1 R^{2 \psi}
\end{equation}
from which we obtain
\begin{align} \label{M2}
\begin{split}
P_i(r) &= r^{\eta + 1/2} (a_1 r^\psi - a_1 R^{2 \psi} r^{-\psi}) \\
&=a_1 R^{\psi} r^{\eta + 1/2} \left[\left(\frac{R}{r}\right)^{- \psi}-  \left(\frac{R}{r}\right)^{\psi}\right]\\
&=a_1 R^{\psi} r^{\eta + 1/2} \left \{ \exp \left[ - \psi\ln\left(\frac{R}{r}\right)\right]- \exp \left[ \psi \ln \left(\frac{R}{r}\right) \right]\right \}\\
& = a_1^* r^{\eta+1/2} \sinh[\psi \ln(R/r)]
\end{split}
\end{align}
where $a_1^* = -2 a_1 R^\psi$. Thus 
\begin{equation}
\frac{dP_i}{dr} = a_1^* r^{\eta-1/2} \{(\eta+1/2) \sinh[\psi \ln(R/r)] - \psi \cosh [ \psi \ln(R/r)]\}
\end{equation}
and hence the first boundary condition implies that
\begin{equation}
\tanh \left[\psi \ln \left( \frac{R}{r_0} \right) \right] = \frac{2 \psi}{1+2\eta}
\end{equation}
This transcendental equation has one unique solutions $\psi_1$ and thus only one small $\lambda_i$ is obtained that satisfies $\Lambda_i < (\eta+ 1/2)^2$. If
\begin{equation} \label{app_nu}
\ln \left(\frac{R}{r_0} \right) \gg \frac{2}{1+2\eta}
\end{equation}
an approximate solution to this equation is: 
\begin{equation} \label{prob_1}
\psi \approx (\eta+1/2)\left[ 1-2\left(\frac{R}{r_0}\right)^{-(1+2\eta)}\right]
\end{equation}
Since $\Lambda_1 = (\eta+1/2)^2 - \psi^2$, we obtain by expanding $\psi$
\begin{align}
\Lambda_1 &= (\eta+1/2)^2 - (\eta+1/2)^2\left[ 1-4\left(\frac{R}{r_0}\right)^{-(1+2\eta)}+ \ldots\right] \\
&\approx (1+2\eta)^2\left(\frac{R}{r_0}\right)^{-(1+2\eta)}
\end{align}
and finally since $\lambda_1^* = \beta \chi^2 \Lambda_1$ and $\eta=5M_A^2/(2 \chi) - 3/2$, we obtain for $\lambda_1^*$
\begin{equation}
\lambda_1^* = \frac{3 \chi^2}{5M_A^2}\left(\frac{5}{\chi}M_A^2-2\right)^2 \left(\frac{r_0}{R}\right)^{5M_A^2/\chi - 2}
\end{equation}
as required.
\subsection*{Case 2: $\Lambda_i \geq (\eta+ 1/2)^2$}
The remaining $\lambda_i^*$s are calculated for $\Lambda_i \geq (\eta+ 1/2)^2$.  Setting $\nu^2 = \Lambda_i  - ( \eta + 1/2 )^2 >0$, the solution to equation \eqref{eqn:Es} is now given by
\begin{equation}
\rho(r) = r^{1/2}(b_1 r^{i \psi} + b_2 r^{- i \psi})
\end{equation}
Following the same procedure as in the previous section, the remaining $P_i$s are calculated as 
\begin{equation} \label{M3}
P_i(r) =  b_1^{*}r^{\eta+1/2} \sin[\nu_i \ln(R/r)]
\end{equation}
where $b_1^{*} = -2ib_1 R^{i \psi}$. Once again, the first boundary conditions results in a transcendental equation, this time of the form
\begin{equation}
\tan \left[\nu \ln \left( \frac{R}{r_0} \right) \right] = \frac{2 \nu}{1+2\eta}
\end{equation}
However, this equation now has an infinite amount of solutions which, if condition \eqref{app_nu} is satisfied, are approximately given by
\begin{equation}
\nu_i \approx (i-1)\pi \left[ 1+\frac{1}{(\eta+1/2)\ln(R/r_0)}\right] \,\,\, i=2,3\dotsc
\end{equation} \label{prob_2}
Therefore, as $ \Lambda_i  = \nu^2  + ( \eta + 1/2 )^2 $, we obtain for the $\Lambda_i$'s
\begin{equation}
\Lambda_i = (i-1)^2\pi^2 \left[ 1+\frac{1}{(\eta+1/2)\ln(R/r_0)}\right]^2 + \left(\eta+\frac{1}{2}\right)^2 \,\, \,i=2,3\dotsc
\end{equation}
and thus for the remaining $\lambda_i^*$s
\begin{multline}
\lambda_i^* = \frac{3 \chi^2}{5 M_A^2}\left\{(i-1)^2\pi^2 \left[ 1+\frac{1}{(5M_A^2/2\chi-1)\ln(R/r_0)}\right]^2 + \left(\frac{5M_A^2}{2\chi} -1\right)^2\right\} \\
i=2,3\dotsc
\end{multline}
as required.

\section{Solving Equation \eqref{or_tran} numerically using the Gauss Seidel Method} \label{app_C}
We begin with the equation we wish to write numerically, namely equation \eqref{or_tran}
\begin{equation} \label{app_parker1}
\frac{\partial f}{\partial t} + V_0 \frac{\partial f}{\partial r} = \frac{2 V_0}{3 r} p \frac{\partial f}{\partial p} + \frac{1}{r^2}\frac{\partial}{\partial r} \left [ r^2 \kappa(r) \frac{\partial f}{\partial r} \right ]   + \frac{1}{p^2} \frac{\partial}{\partial p} \left [ p^2 D(p,r) \frac{\partial f}{\partial p} \right ] +Q - \frac{f}{\tau_L(r)}
\end{equation} 
We recast the variables into dimensionless quantities, as follows
\begin{align}
\tilde{r} = \frac{r}{r_\text{min}} && \tilde{p} =\frac{p}{p_I} &&  \tilde{y}=\ln {\tilde{p}}
\end{align}
where the normalising values have their previous meanings. Thus, the steady state form of equation \eqref{app_parker1} is now given by
\begin{multline} \label{transeq}
\frac{V_0}{r_\text{min}} \frac{\partial f}{\partial \tilde{r}} = \frac{2 V_0}{3 r_\text{min} \tilde{r}} \frac{\partial f}{\partial \tilde{y}} + \frac{1}{r_\text{min}^2 \tilde{r}^2} \frac{\partial}{\partial \tilde{r}} \left [\tilde{r}^2 \kappa(\tilde{r})\frac{\partial f}{\partial \tilde{r}} \right ]  
+e^{-3\tilde{y}} \frac{\partial}{\partial \tilde{y}} \left [ D(\tilde{y},\tilde{r}) e^{\tilde{y}} \frac{\partial f}{\partial \tilde{y}} \right ] +Q -\frac{f}{\tau_L(\tilde{r})} 
\end{multline}
Multiplying across by $r_\text{min}/V_0$, we obtain
\begin{multline}  \label{fin1}
\frac{\partial f}{\partial \tilde{r}} = \frac{2}{3 \tilde{r}} \frac{\partial f}{\partial \tilde{y}} + \frac{1}{r_\text{min} V_0 \tilde{r}^2} \frac{\partial}{\partial \tilde{r}} \left [\tilde{r}^2 \kappa(\tilde{r})\frac{\partial f}{\partial \tilde{r}} \right ]  
+\frac{r_\text{min}e^{-3\tilde{y}}}{V_0} \frac{\partial}{\partial \tilde{y}} \left [ D(\tilde{y},\tilde{r}) e^{\tilde{y}} \frac{\partial f}{\partial \tilde{y}} \right ] +\frac{r_\text{min}Q}{V_0} -\frac{r_\text{min}}{V_0\tau_L(\tilde{r})}f 
\end{multline}
We now approximate these derivatives by using a finite difference grid. In order to ensure that the solutions are accurate, we adopt the following second order finite difference approximations
\begin{equation}  
\frac{\partial f}{\partial \tilde{r}} = \frac{-f_{i+2j} + 8f_{i+1j} - 8f_{i-1 j} + f_{i-2j}}{12 \Delta \tilde{r}} 
\end{equation}
\begin{equation}  
\frac{\partial f}{\partial \tilde{y}} =  \frac{-f_{ij+2} + 8f_{ij+1}-8f_{ij-1} + f_{ij-2}}{12 \Delta \tilde{y}} 
\end{equation}
\begin{multline}  
\frac{\partial}{\partial \tilde{r}} \left [\tilde{r}^2 \kappa(\tilde{r})\frac{\partial f}{\partial \tilde{r}} \right ]  =\frac{1}{12 (\Delta \tilde{r})^2}\left[(f_{i+1j} - f_{i+2j}) \tilde{r}_{i+3/2}^2 \kappa_{i+3/2} \right. \\
\left. +15(f_{i+1j}- f_{ij}) \tilde{r}_{i+1/2}^2 \kappa_{i+1/2}  - 15(f_{ij}- f_{i-1j}) \tilde{r}_{i-1/2}^2 \kappa_{i-1/2} \right. \\
\left. + (f_{i-1j}- f_{i-2j}) \tilde{r}_{i-3/2}^2 \kappa_{i-3/2}\right] 
\end{multline}
\begin{multline}  
\frac{\partial}{\partial \tilde{y}} \left [ D(\tilde{y},\tilde{r}) e^{\tilde{y}} \frac{\partial f}{\partial \tilde{y}} \right ] =  \frac{1}{12 (\Delta \tilde{y})^2} \left[(f_{ij+1}- f_{ij+2}) D_{ij+3/2} e^{ \tilde{y}_{j+3/2}} + 15(f_{ij+1}- f_{ij}) D_{ij+1/2} e^{ \tilde{y}_{j+1/2}} \right.\\
\left. - 15(f_{ij}- f_{ij-1}) D_{ij-1/2} e^{ \tilde{y}_{j-1/2}} + (f_{ij-1}- f_{ij-2}) D_{ij-3/2} e^{ \tilde{y}_{j-3/2}} \right]
\end{multline}
where the $i$th and $j$th indices refer to space and momentum respectively. Inserting each of these approximations into equation \eqref{fin1} and rearranging, we obtain an equation of the form
\begin{multline}
f_{ij} = \frac{\Xi_{i}}{\alpha_{ij}}f_{i-2j} + \frac{\beta_{i}}{\alpha_{ij}}f_{i-1j} + \frac{\delta_{i}}{\alpha_{ij}}f_{i+1j} + \frac{\Theta_{i}}{\alpha_{ij}}f_{i+2j} + \frac{\psi_{i}}{\alpha_{ij}}f_{ij-2} + \frac{\gamma_{ij}}{\alpha_{ij}}f_{ij-1} \\
+ \frac{\Sigma_{ij}}{\alpha_{ij}}f_{ij+1} + \frac{\Psi_{i}}{\alpha_{ij}}f_{ij+2} +  \frac{r_\text{min}Q}{V_0 \alpha_{ij}}
\end{multline}
where these spatial and momentum dependent quantities are defined as
\begin{multline}
\alpha_{ij} = \frac{15}{12r_\text{min} V_0 \tilde{r}_i^2 (\Delta \tilde{r})^2}( \tilde{r}_{i+1/2}^2 \kappa_{i+1/2}+ \tilde{r}_{i-1/2}^2 \kappa_{i-1/2})\\
+ \dfrac{15r_\text{min} e^{-3 \tilde{y}_j}}{12 V_0 (\Delta \tilde{y})^2} (D_{j+1/2}e^{ \tilde{y}_{j+1/2}} + D_{j-1/2}e^{ \tilde{y}_{j-1/2}}) + \frac{r_\text{min}}{V_0 \tau_{Li}}
\end{multline}
\begin{align}
\Xi_i =   - \frac{1}{12 \Delta \tilde{r}}
- \dfrac{1}{12r_\text{min} V_0 \tilde{r}_i^2 (\Delta \tilde{r})^2} \tilde{r}_{i-3/2}^2 \kappa_{i-3/2}
\end{align}
\begin{align}
\begin{split}
\beta_i = \frac{8}{12 \Delta \tilde{r}} +  \frac{15}{12r_\text{min} V_0 \tilde{r}_i^2 (\Delta \tilde{r})^2} \tilde{r}_{i-1/2}^2 \kappa_{i-1/2} + \frac{1}{12r_\text{min} V_0 \tilde{r}_i^2 (\Delta \tilde{r})^2} \tilde{r}_{i-3/2}^2 \kappa_{i-3/2} 
\end{split}
\end{align}
\begin{align}
\delta_i =  - \frac{8}{12 \Delta \tilde{r}} +  \frac{15}{12r_\text{min} V_0 \tilde{r}_i^2 (\Delta \tilde{r})^2} \tilde{r}_{i+1/2}^2 \kappa_{i+1/2} +  \frac{1}{12r_\text{min} V_0 \tilde{r}_i^2 (\Delta \tilde{r})^2} \tilde{r}_{i+3/2}^2 \kappa_{i+3/2} 
\end{align}
\begin{align}
\Theta_i =  \frac{1}{12 \Delta \tilde{r}} - \frac{1}{12r_\text{min} V_0 \tilde{r}_i^2 (\Delta \tilde{r})^2} \tilde{r}_{i+3/2}^2 \kappa_{i+3/2}
\end{align}
\begin{align}
\psi_{ij} = \frac{2}{36 \tilde{r}_i \Delta \tilde{y}} - \dfrac{r_\text{min} e^{-3 \tilde{y}_j}}{12 V_0 (\Delta \tilde{y})^2} D_{ij-3/2} e^{ \tilde{y}_{j-3/2 }}
\end{align}
\begin{align}
\gamma_{ij} =-\frac{16}{36 \tilde{r}_i \Delta \tilde{y}} + \dfrac{15r_\text{min} e^{-3 \tilde{y}_j}}{12 V_0 (\Delta \tilde{y})^2} D_{ij-1/2} e^{ \tilde{y}_{j-1/2 }} +\dfrac{r_\text{min} e^{-3 \tilde{y}_j}}{12 V_0 (\Delta \tilde{y})^2} D_{ij-3/2} e^{ \tilde{y}_{j-3/2}}
\end{align}
\begin{align}
\Sigma_{ij} =\frac{16}{36 \tilde{r}_i \Delta \tilde{y}} +\dfrac{15r_\text{min} e^{-3 \tilde{y}_j}}{12 V_0 (\Delta \tilde{y})^2} D_{ij+1/2} e^{ \tilde{y}_{j+1/2 }} +  \dfrac{r_\text{min} e^{-3 \tilde{y}_j}}{12 V_0 (\Delta \tilde{y})^2} D_{ij+3/2} e^{ \tilde{y}_{j+3/2 }}
\end{align}
\begin{align}
\Psi_{ij}  &=-\frac{2}{36 \tilde{r}_i \Delta \tilde{y}} - \dfrac{r_\text{min} e^{-3 \tilde{y}_j}}{12 V_0 (\Delta \tilde{y})^2} D_{ij+3/2} e^{ \tilde{y}_{j+3/2 }} 
\end{align}
This finite difference scheme is solved using the Gauss Seidel method, a scheme that is commonly used to numerically solve steady-state differential equations. We begin with an initial guess of the distribution, namely $f_{ij}^0$. Then, we use the following equation to calculate better estimates at each $k$th attempt semi-implicitly
\begin{multline} \label{gs_num}
f_{ij}^{k+1} = \frac{\Xi_{i}}{\alpha_{ij}}f_{i-2j}^{k+1} + \frac{\beta_{i}}{\alpha_{ij}}f_{i-1j}^{k+1} + \frac{\delta_{i}}{\alpha_{ij}}f_{i+1j}^k + \frac{\Theta_{i}}{\alpha_{ij}}f_{i+2j}^{k} + \frac{\psi_{i}}{\alpha_{ij}}f_{ij-2}^{k+1}  + \frac{\gamma_{ij}}{\alpha_{ij}}f_{ij-1}^{k+1} \\
+ \frac{\Sigma_{ij}}{\alpha_{ij}}f_{ij+1}^k + \frac{\Psi_{i}}{\alpha_{ij}}f_{ij+2}^k + \tau_{C0}\frac{Q}{\alpha_{ij}}
\end{multline}
We continue to evolve the distribution to more accurate solutions until a predefined stopping criteria is obtained.
\vspace{10mm}

\bibliography{references}
\end{document}